\documentclass{elsart}

\usepackage{graphicx}
\usepackage{amsmath}

\begin{document}

\begin{frontmatter}


\begin{flushright}
RIKEN-TH-110
\end{flushright}

\title{Partially quenched chiral perturbation theory in the $\epsilon$-regime}


\author[1]{Poul~H.~Damgaard}
\address[1]{The Niels Bohr Institute, The Niels Bohr International Academy,
Blegdamsvej 17 DK-2100 Copenhagen {\O}
Denmark}
\ead{phdamg@nbi.dk}

\author[2]{Hidenori~Fukaya}
\address[2]{
  Theoretical Physics Laboratory, RIKEN,
  Wako 351-0198, Japan
}
\ead{hfukaya@riken.jp}

\begin{abstract}
We calculate meson correlators 
in the $\epsilon$-regime  within partially quenched 
chiral perturbation theory.
The valence quark masses and sea quark masses can be
chosen arbitrary and all non-degenerate.
Taking some of the sea quark masses to infinity, one
obtains a smooth connection among the theories with
different number of flavors, as well as the quenched
theory.
These results can be directly compared with lattice QCD
simulations. 
\end{abstract}

%\begin{keyword}
%\end{keyword}

\end{frontmatter}
\renewcommand{\theequation}{\arabic{section}.\arabic{equation}}

%%%%%%%%%%%%%%%%%%%%%%%%%%%
%%%%%%%%%%%%%%%%%%%%%%%%%%%
\section{Introduction}
%%%%%%%%%%%%%%%%%%%%%%%%%%%
%%%%%%%%%%%%%%%%%%%%%%%%%%%

In the low energy limit, the dynamics of QCD 
is described by the pion fields which appear as 
pseudo--Nambu-Goldstone bosons accompanying
the spontaneous breaking of chiral symmetry.
Chiral dynamics and chiral perturbation theory (ChPT),
play an essential role in understanding 
the interactions among the pions themselves,
as well as their couplings
with the other (heavier) hadrons and sources.

The fundamental parameters of ChPT
%the low-energy constants of QCD, 
are unknown coupling constants in the effective
theory. But they can be
determined by non-perturbative
and first-principle calculations in the underlying 
theory, QCD. The most important low-energy constants
are the chiral condensate $\Sigma$, and the pion
decay constant, $F$, both at leading order.
%, and many others at higher order in ChPT.
Numerical simulations of lattice QCD 
give currently the most promising approach to 
achieving this non-perturbative determination. 
Because such computer simulations
necessarily are restricted to finite volumes,
it is important to understand finite-volume
effects in the effective field theory.

Near the chiral limit,
the finite-volume effects become 
increasingly significant due to the diverging 
correlation length of the Goldstone bosons.
In particular, when the pion correlation length, or
the inverse of the pion mass $m_\pi$ overcomes 
the size of the box $L$, 
%%%%%%%%%% Eq e-regime definition %%%%%%%%%%%%%
\begin{equation}
\frac{1}{\Lambda_{\rm QCD}}\ll L \ll \frac{1}{m_\pi},
\end{equation}
where $\Lambda_{\rm QCD}$ is the QCD scale,
the zero-momentum mode has to be 
treated non-perturbatively and
the ChPT has to be performed in a way that achieves this
in a systematic fashion:
the so-called $\epsilon$-expansion
\cite{Gasser:1987ah, Neuberger:1987zz, Hansen:1990un, 
%Hansen:1990yg, 
Hasenfratz:1989pk}.
New counting rules are needed to order the perturbative
expansion in this case. In units of the ultraviolet cut-off,
the expansion parameter can be defined as $\epsilon$, where,
unusually, the pion mass $m_{\pi}$ is not treated as being of the same
order as pion momentum $p$. Instead, 
%%%%%%%%%% Eq e-expansion %%%%%%%%%%%%%%%%%%%%%
\begin{equation}
m_\pi ~\sim~ 
 p^2  ~\sim~ \epsilon^2 ,
\end{equation}
while the inverse of space-time volume $V$ and quark mass $m_q$ 
are being of ${\cal O}(\epsilon^4)$. The particularly important
combination $m_q \Sigma V$ is thus treated as of order unity.
With this expansion, the precise analytical predictions
for physical observables in the low-energy sector of QCD at 
finite volume $V$ are expressed in terms of the 
low-energy constants at infinite volume. By comparing
numerical results at finite volume with these predictions, 
one can thus extract the infinite-volume constants directly from
finite-volume simulations, without the need for extrapolations
of data to infinite volume. The closer one gets to the chiral 
limit, the bigger is the advantage of this approach.

A few years ago, the predictions for correlation functions 
of ChPT were extended to the cases of both quenched QCD and 
full QCD at sectors of fixed gauge-field topology
\cite{Damgaard:2001js, Damgaard:2002qe}. 
For the chiral condensate, the studies were also extended to
the partially quenched cases \cite{Damgaard:1999ic,Damgaard:2000di}.
%, which is described by the expansion of the valence quark masses.
In quenched QCD
these analytical predictions suffer in the $\epsilon$-regime
from quenched finite-volume logarithms \cite{Damgaard:2001xr}.
Strictly speaking, such logarithms prevent taking the
infinite-volume limit, and basically invalidate the whole
chiral expansion in this regime for the quenched theory.
The hope is that in finite ranges of volume, the resulting
predictions may still have a certain range of validity.
For both quenched and unquenched cases these computations
were restricted to the case of degenerate, light, quarks.

As mentioned above,
the great advantage of the predictions for correlators
in the $\epsilon$-regime is that they
appear almost tailored for numerical lattice computations
near the chiral limit. Indeed, exploratory studies of these
correlation functions have already shown the great potential 
\cite{Giusti:2002sm, Bietenholz:2003bj,
Giusti:2003iq, Giusti:2004yp, Ogawa:2005jn, Fukaya:2005yg,
Bietenholz:2006fj}. In particular, the possibility of using
to one's advantage the role played by fixing topology
%topological charge 
in finite volume has been clearly demonstrated.
Also the analytic handle one has on the quark mass dependence 
due to the finite size effects in the $\epsilon$-regime
has proved helpful in 
reducing the systematic errors of lattice simulations.
%Moreover, it is known that the dynamics of the 
%lowest Dirac eigenmodes are equivalent to that of the zero modes
%of ChPT in the $\epsilon$-regime via chiral Random Matrix Theory, 
%%% refs RMT
%which provides another promising method to investigate 
%the low-energy dynamics of QCD.

In this paper, we generalize these analytical computations of
correlation functions in the $\epsilon$-regime to
the partially quenched theory
with separate valence and sea quarks, 
both of which are taken to be non-degenerate. 
The chiral condensate and pseudo-scalar and scalar meson correlators
are calculated as functions of non-degenerate quark masses,
topological charge, and the volume of the Euclidean space-time $V$.
In a separate forthcoming publication \cite{Damgaard:preparation} 
%{\bf (ref!)} 
we will
present the analogous results for the vector and axial vector 
channels.

In practice, the $\epsilon$-regime is not trivially reached in numerical 
simulations. It is therefore important to be able to go close to the chiral
limit, but still only marginally in the $\epsilon$-regime, while
valence quark masses are taken to that regime. In fact, although
our aim in this paper is the $\epsilon$-regime,
our partially quenched chiral perturbation theory (PQChPT)
\cite{Bernard:1993sv, Sharpe:2001fh}
in the $\epsilon$-regime
smoothly connects all the theories 
with a different number of flavors 
as a function of the sea quark masses. In this way, our
calculation interpolates between the $\epsilon$-regime
and the more conventional $p$-regime, and in
one kinematical regime also mixes the two expansions. For this
reason we provide expressions not just with the (simpler)
$\epsilon$-expansion propagators, but the more general expressions.
If one of the sea quark masses is taken to infinity in the 
$N_f$-flavor theory, it converges to the $(N_f-1)$-flavor theory.
Even the quenched theory can be obtained 
by carefully introducing the flavor singlet field before 
taking all the sea quark masses to infinity. 
%When the sea and valence quark masses separate in scale well-known
%quenching artifacts show up \cite{Damgaard:2001js}\footnote{We thank
%P. Hernandez for stressing this point.}, an issue that is
%discussed in Appendix B.
Of course, the
low-energy constants in addition have an inherent flavor dependence
that is beyond our control.

Our results have wide applicability to unquenched
lattice QCD studies near the chiral limit 
\cite{Fukaya:2006xp, Fukaya:2007fb, Fukaya:2007yv}.
One can choose various valence quark masses 
with a fixed sea quark mass.
The partially quenched condensate and
meson correlators can be compared
with simulations of heavier quarks which are perhaps
only marginally in the $\epsilon$-regime (or beyond)
while the 
valence quark mass is still in the $\epsilon$-regime.

This paper is organized as follows. 
In Sec.\ref{sec:partitionfunction}, 
we describe the leading contribution of
the partition function of PQChPT
in the $\epsilon$-expansion.
We discuss, in particular, the exact non-perturbative
integral of the zero-modes \cite{Splittorff:2002eb,Fyodorov:2002wq} 
which plays a crucial role in deriving both the chiral condensate
and meson correlators in this extended theory.
As one fundamental building block of this work, 
the chiral condensate and its 1-loop level correction
are obtained in Sec.\ref{sec:condensate}.
The exact zero-mode integrals in the replica limit
are calculated in Sec.\ref{sec:zero-modes}. 
We also derive a non-trivial identity which follows from
the unitarity of the group integrals.
In Sec.\ref{sec:meson}, our main results on meson correlators
are presented. We plot $N_f=2$ connected pseudo-scalar and scalar
correlators as examples.
The conclusions are given in Sec.\ref{sec:conclusion}.

%%%%%%%%%%%%%%%%%%%%%%%%%%%%%%%%%%%%%%%%%%%%%%%
%%%%%%%%%%%%%%%%%%%%%%%%%%%%%%%%%%%%%%%%%%%%%%%
\section{The partition function of PQChPT}
\label{sec:partitionfunction}
\setcounter{equation}{0}
%%%%%%%%%%%%%%%%%%%%%%%%%%%%%%%%%%%%%%%%%%%%%%%
%%%%%%%%%%%%%%%%%%%%%%%%%%%%%%%%%%%%%%%%%%%%%%%

Our starting point is the $(N_f+N)$-flavor chiral Lagrangian,
%%%%%%%%%%%%%%% Eq ChPT Lagrangian %%%%%%%%%%%%%%%
\begin{eqnarray}
\mathcal{L}
&=&
\frac{F^2}{4}{\rm Tr}
(\partial_\mu U(x)^\dagger\partial_\mu U(x) )
-\frac{\Sigma}{2}{\rm Tr}
(\mathcal{M}^{\dagger}U(x)+\mathcal{M}U(x)^\dagger)
\nonumber\\
&&
+\frac{m_0^2}{2N_c}\Phi_0^2(x)
+\frac{\alpha}{2N_c}
\partial_\mu \Phi_0(x)\partial_\mu \Phi_0(x),
\end{eqnarray}
where $\Sigma$ and $F$ are the chiral condensate and the pion 
decay constant at infinite volume, both in the chiral limit.
In the mass matrix 
%%%%%%%%%%%%%% Eq mass matrix %%%%%%%%%%%%%%%%%
\begin{equation}
\mathcal{M}={\rm diag}(\underbrace{m_1,m_2,\cdots, m_{N_f}}_{N_f}, 
\underbrace{m_v,\cdots, m_v}_{N}),
\end{equation}
we have in mind a situation in which the valence quark mass
is always taken in the $\epsilon$-regime of $m_v\Sigma V
\sim {\cal O}(1)$,
while the physical sea quark mass $m_i$ may vary more freely.
Unlike standard chiral perturbation theory,
$U(x)$ is an element of the $U(N_f+N)$ group, and the
flavor-singlet field,
$\Phi_0(x) \equiv \frac{-iF}{\sqrt{2}}{\rm Tr}\ln U(x)$,
is introduced explicitly as a physical degree of freedom 
with additional constants, $m_0$ and $\alpha$.
The number of colors is denoted by $N_c$. As is well known,
in this partially quenched theory 
one can normally take the $m_0\to \infty$ limit 
without difficulty. In terms of first replicated and then
quenched valence quarks one is then going from
$U(N_f+N)$ to $SU(N_f)$ in a smooth way. Then $\Phi_0$ 
can be decoupled from the theory. Of course, trouble arises
again if we consider the theory in a regime where the
sea quark masses $m_i$ have effectively decoupled. We will
discuss this issue below.
% when some of the sea quark masses $m_i$'s are small enough to
% produce $U(1)$ chiral anomaly effect to give large mass $m_0$
% to the singlet field.

Separating the zero-mode, $U_0$,
and the non-zero modes, $\xi(x)$, 
\begin{equation}  
U(x) ~=~ U_0 \exp(i\sqrt{2}\xi(x)/F) ~,
\end{equation}
we consider three types of expansion of the partition function
in a sector of fixed topological charge $\nu$:
\begin{enumerate}
\item Both of the valence and sea quarks are in the $\epsilon$-regime:
%%%%%%%%%%%%%% Eq partition function e-regime v s %%%%%%%%%%
\begin{eqnarray}
\label{eq:Ze+e}
\hspace{-.4in}
Z^\nu_{N_f+N}(m_v,\{m_i \})
&=&
\int_{U(N_f+N)}dU_0  d\xi \det U_0^\nu
%\nonumber\\
%&&
%&&\hspace{-1in}
\exp \left[\frac{\Sigma V}{2}
{\rm Tr}[ \mathcal{M}^{\dagger}U_0+\mathcal{M}U_0^\dagger]\right.
\nonumber\\
&&\hspace{.2in}
\left.+\int d^4 x
\left(
-\frac{1}{2}{\rm Tr}[\partial_\mu \xi \partial_\mu \xi]
%-\frac{m^2_0}{2N_c}({\rm Tr }\xi)^2
%-\frac{\alpha}{2N_c}(\partial_\mu {\rm Tr}\xi)^2
\right)+\cdots
\right],
%\nonumber\\
\end{eqnarray}
where the $m_0\to \infty$ limit is taken and the singlet $\Phi_0$
is decoupled from the theory. 

\item The sea quarks are in the $p$-regime 
(but light enough that we can still 
disregard effects of the singlet field):
%%%%%%%%%%%%%% Eq partition function e-regime v + p-regime s %%%%%%%%%%
\begin{eqnarray}
\label{eq:Ze+p}
\hspace{-.4in}
Z^\nu_{N_f+N}(m_v,\{m_i \})
&=&
\int_{U(N_f+N)}dU_0  d\xi \det U_0^\nu
%\nonumber\\
%&&\hspace{-1in}
\exp \left[\frac{\Sigma V}{2}
{\rm Tr}[ \mathcal{M}^{\dagger}U_0+\mathcal{M}U_0^\dagger]\right.
\nonumber\\
&&\hspace{-1in}
\left.+\int d^4 x
\left(
-\frac{1}{2}{\rm Tr}[\partial_\mu \xi \partial_\mu \xi]
%-\frac{m^2_0}{2N_c}({\rm Tr }\xi)^2
%-\frac{\alpha}{2N_c}(\partial_\mu {\rm Tr}\xi)^2
%\right)
%\right.
%\nonumber\\
%&&\left.
-\left(\frac{\Sigma}{F^2}\right){\rm Tr}[\mathcal{M}\xi^2]
+\cdots
\right)
\right] ~.
%\nonumber\\
\end{eqnarray}
Here the valence sector is expanded as in $\epsilon$-expansion,
while the sea sector is considered in the $p$-regime.
Note that the mass term in the valence sector,
$m_v\Sigma \xi^2/F^2$, is of ${\cal O}(\epsilon^4)$ 
but not ignored here, in order to see a smooth transition
to the $p$-regime.

\item The sea quarks are heavy:
%%%%%%%%%%%%%% Eq partition function quenched limit %%%%%%%%%%
\begin{eqnarray}
\label{eq:Zquench}
\hspace{-0.4in}
Z^\nu_{N_f+N}(m_v,\{m_i \})
&=&
\int_{U(N_f+N)}dU_0  d\xi \det U_0^\nu
\exp \left[\frac{\Sigma V}{2}
{\rm Tr}[ \mathcal{M}^{\dagger}U_0+\mathcal{M}U_0^\dagger]\right.
\nonumber\\
&&%\hspace{-1.2in}
%\left.
+\int d^4 x
\left(
-\frac{1}{2}{\rm Tr}[\partial_\mu \xi \partial_\mu \xi]
-\left(\frac{\Sigma}{F^2}\right){\rm Tr}[\mathcal{M}\xi^2]
\right.
\nonumber\\
&&
\left.\left.
-\frac{m^2_0}{2N_c}({\rm Tr }\xi)^2
-\frac{\alpha}{2N_c}(\partial_\mu {\rm Tr}\xi)^2
\right)+\cdots
\right].
\end{eqnarray}
\end{enumerate}
As the sea quark mass increases,
new terms (the non-zero mode's mass term, 
and the singlet fields) come in. 
Thus, Eq.(\ref{eq:Zquench}) is the most general form.

%As remarked in the Introduction,
%one should keep in mind that quenching artifacts begin to
%show up when one separates the scale between sea and 
%valence quarks. In particular, chiral divergences in the
%valence quark masses invalidate the expansion for large
%topological charge $\nu$ as one takes the chiral limit
%in the valence sector while keeping sea quark masses
%fixed. This is discussed in more detail in 
%Appendix \ref{app:nulimit}.
In the following, %whenever the two scales separate largely,
we implicitly restrict ourselves to sectors of small enough fixed
topology $\nu$ for the expansion to be valid \cite{Bernardoni:2007hi}. 
Indeed, this is an essential assumption 
%that ${\rm Tr}[\mathcal{M}(1-(U_0+U_0^\dagger)/2)\xi^2]$ 
%remains small 
so that
%is small in the both regime; ${\cal O}(\epsilon^4)$ in 
%the $\epsilon$-regime, and $U_0$ is close to 1 in the $p$-regime,
in Eq.(\ref{eq:Ze+p}) and Eq.(\ref{eq:Zquench}) one needs only 
the mass term ${\rm Tr}[\mathcal{M}\xi^2]$
instead of ${\rm Tr}[\mathcal{M}(U_0+U_0)\xi^2]/2$ 
\footnote{We thank P. Hernandez for stressing this point.}.

%As the sea quark mass increases, 
%and effectively quenches
%on the scale of the valence quark mass, the singlet field
%can no longer be ignored. 
%Thus, Eq.(\ref{eq:Zquench}) is the most general form.

For the zero-mode integrals, one needs exact formulae of the
group integrals over $U(N_f+N)$ and means of taking 
the replica limit. As described in detail in ref.
\cite{Damgaard:2001js}, if we wish to consider correlation
functions with $N_v$ external valence quarks we must embed the
$N_v$ valence quarks in a theory with $N$ replicated quarks
in total (of which $N - N_v$ do not couple to the external sources), 
and then take the limit $N \to 0$. 
Alternatively, one can consider a theory with 
$N_v$ additional bosonic flavors of common mass $m_v$. 
It is easy to understand in the quark determinant picture
that this limit is equivalent to the replica limit:
%%%%%%%%%%%%% Eq 2 replica methods equivalence %%%%%%%%%%%% 
\begin{eqnarray}
\lim_{N \to 0} \det (D+m)^{N_f}\left(\prod_i^{N_v}\det (D+m_v+J_i)\right)
\det(D+m_v)^{N-N_v}
\nonumber\\
= \det (D+m)^{N_f}
\frac{\prod_i^{N_v}\det (D+m_v+J_i)}{\det (D+m_v)^{N_v}}.
\end{eqnarray}
Here the left hand side is the replica prescription, while
the right hand side is the prescription of the graded formalism.

The zero-mode partition function of $n$ bosons and $m$ fermions
are analytically known 
\cite{Splittorff:2002eb, Fyodorov:2002wq},
%%%%%%%%%%%%% Eq zero-mode partition function %%%%%%%%%%%%%
\begin{equation}
\mathcal{Z}^\nu_{n,m}(\{\mu_i\})
=
\frac{\det[\mu_i^{j-1}\mathcal{J}_{\nu +j-1}(\mu_i)]_{i,j=1,\cdots n+m}}
{\prod_{j>i=1}^n(\mu_j^2-\mu_i^2)\prod_{j>i=n+1}^{n+m}(\mu_j^2-\mu_i^2)},
\end{equation}
where $\mu_i=m_i\Sigma V$.
Here $\mathcal{J}$'s are defined as
$\mathcal{J}_{\nu+j-1}(\mu_i)\equiv (-1)^{j-1} K_{\nu+j-1}(\mu_i)$ 
for $i=1,\cdots n$ and 
$\mathcal{J}_{\nu+j-1}(\mu_i)\equiv I_{\nu+j-1}(\mu_i)$ 
for $i=n+1,\cdots n+m$, 
where $I_\nu$ and $K_\nu$ are the modified Bessel functions.
Of particular importance is the case $(n, m)=(1, N_f+1)$:
%%%%%%%%%%%%%% Eq (n, m)=(1, N_f+1) partition function %%%%%%%%%%
\begin{eqnarray}
\label{eq:zero-mode}
\mathcal{Z}^\nu_{1,1+N_f}(x | y,\{z_i\})
&=&
\frac{1}{\prod_{i=1}^{N_f}(z_i^2-y^2)
\prod_{k>j}^{N_f}(z_k^2-z_j^2)}
\nonumber\\
&&\hspace{-1in}
\times \det \left(
\begin{array}{ccccc}
K_\nu(x) & I_\nu(y) & I_\nu(z_1) & I_\nu(z_2) & \cdots\\
-x K_{\nu+1}(x) & yI_{\nu+1}(y) & z_1I_{\nu+1}(z_1) 
& z_2I_{\nu+1}(z_2) & \cdots\\
x^2 K_{\nu+2}(x) & y^2I_{\nu+2}(y) & z_1^2I_{\nu+2}(z_1) 
& z^2_2I_{\nu+2}(z_2) & \cdots\\
\cdots & \cdots & \cdots & \cdots & \cdots
\end{array}
\right),
\end{eqnarray}
where $x=m_b\Sigma V$ ($m_b$ denotes the bosonic quark mass), 
$y=m_v\Sigma V$ and $z_i=m_i\Sigma V$.
One notes that 
%\footnote{Eq.(\ref{eq:xylimit}) should be always true, 
%but the analytical proof has not been known except 
%for small $N_f$ cases where one can explicitly calculate it.}
%%%%%%%%%%%%%%% Eq x->y    limit %%%%%%%%%%%%%%%%
\begin{equation}
\label{eq:xylimit}
\lim_{x\to y}\mathcal{Z}^\nu_{1,1+N_f}(x | y,\{z_i\})=
\mathcal{Z}^\nu_{0,N_f}(\{z_i\}),
\end{equation}
and therefore,
%%%%%%%%%%%%%%% Eq x, y derivative %%%%%%%%%%%%%%
\begin{equation}
-\lim_{x\to y}\partial_x
\mathcal{Z}^\nu_{1,1+N_f}(x | y, \{z_i\})=
\lim_{x\to y}
\partial_y \mathcal{Z}^\nu_{1,1+N_f}(x | y, \{z_i\}).
\end{equation}
It is also remarkable that
%%%%%%%%%%%%%%%% Eq Z N_f -> N_f-1 %%%%%%%%%%%%%%%%%
\begin{eqnarray}
\label{eq:Qlim}
\mathcal{Z}^\nu_{1,1+N_f}(x | y, \{z_1,z_2,\cdots,
z_{j-1},z_j\to \infty, z_{j+1},\cdots ,z_{N_f} \})=
\hspace{0.5in}
\nonumber\\
\mathcal{Z}^\nu_{1,1+(N_f-1)}(x | y, \{z_1,z_2,\cdots,
z_{j-1}, z_{j+1},\cdots ,z_{N_f} \})
\nonumber\\
\times 
\left.\left(\left[z_j^{1-N_f}\mathcal{Z}^{\nu+N_f+1}_{0,1}(z_j)
\right]\right|_{z_j\to \infty}\right),
\end{eqnarray}
which is consistent with the intuitive notion that the 
$N_f-1$-flavor theory
can be obtained in the limit of large sea-quark mass, $m_j\to \infty$,
up to a normalization
factor $\left.\left[z_j^{1-N_f}\mathcal{Z}^{\nu+N_f+1}_{0,1}(z_j)
\right]\right|_{z_j\to \infty}$. %{\bf isn't it unity??}.
This decoupling is of course more general. For example,
in the case of $\{z_i\to \infty$ (for all $i$)$\}$ 
one obtains $\mathcal{Z}^\nu_{1,1}$,
the leading partition function of
quenched chiral perturbation theory.

For different valence quarks (with non-degenerate valence masses), 
we will also need
$\mathcal{Z}_{2,2+N_f}$:
%%%%%%%%%%%%%%%% Eq (2,2+N_f) Z %%%%%%%%%%%%%%%%%%%%%%%%%%%
\begin{eqnarray}
\label{eq:zero-mode-2val}
\mathcal{Z}^\nu_{2,2+N_f}(x_1,x_2|y_1,y_2,\{z_i\})
=\hspace{2.2in}\nonumber\\
\frac{1}{(x_2^2-x_1^2)(y^2_2-y^2_1)\prod_{i=1}^{N_f}(z_i^2-y_2^2)(z_i^2-y_1^2)
\prod_{k>j}^{N_f}(z_k^2-z_j^2)}
\hspace{.5in}
\nonumber\\
\hspace{-.3in} 
\times \det \left(
\begin{array}{cccccc}
K_\nu(x_1) & K_\nu(x_2) & I_\nu(y_1) &I_\nu(y_2) 
& I_\nu(z_1) %& I_\nu(z_2) 
& \cdots\\
-x_1 K_{\nu+1}(x_1) &-x_2 K_{\nu+1}(x_2) & y_1I_{\nu+1}(y_1) &
y_2I_{\nu+1}(y_2)
& z_1I_{\nu+1}(z_1) %& z_2I_{\nu+1}(z_2) 
& \cdots\\
x_1^2 K_{\nu+2}(x_1) &x_2^2 K_{\nu+2}(x_2) 
& y_1^2I_{\nu+2}(y_1) & y_2^2I_{\nu+2}(y_2) & z_1^2I_{\nu+2}(z_1) 
%& z^2_2I_{\nu+2}(z_2) 
& \cdots\\
\cdots & \cdots & \cdots & \cdots 
& \cdots
\end{array}
\right).\nonumber\\
\end{eqnarray}

Let us now define the propagator  \cite{Damgaard:2000gh} of
the fluctuation field $\xi$:
%%%%%%%%%%%%% propagator definition %%%%%%%%%%%%%%%%%%%%%%%%
\begin{eqnarray}
\bar{P}_{(ij)(kl)}(x-y)&\equiv& \langle \xi_{ij}(x)\xi_{kl}(y) \rangle
\nonumber\\
&=& \left\{
\begin{array}{lc}
\delta_{il}\delta_{jk}\bar{\Delta}(M^2_{i j}|x-y) &(i\neq j)\\
\delta_{ik}\delta_{kl}\bar{\Delta}(M^2_{i i}|x-y)-
\delta_{kl}\bar{G}(M^2_{ii},M^2_{kk}|x-y)&(i=j)
\end{array}
\right. ,\nonumber\\
\end{eqnarray}
where the indices $i,j\cdots$ can be taken both in the valence and
sea sectors. Here $M^2_{i j}=(m_i+m_j)\Sigma/F^2$ and
%%%%%%%%%%%%% Feynmann rules %%%%%%%%%%%%%%%%%%%%%%%%%%%%%%%%
\begin{eqnarray}
\bar{\Delta}(M^2_{i j}|x) &\equiv& \frac{1}{V}\sum_{p\neq 0}
\frac{e^{ipx}}{p^2+M^2_{ij}},\\
\bar{G}(M^2_{i i},M^2_{j j}|x) &\equiv& \frac{1}{V}\sum_{p\neq 0}
\frac{e^{ipx}(m_0^2+\alpha p^2)/N_c}{(p^2+M^2_{ii})(p^2+M^2_{jj})
\mathcal{F}(p^2)},\\
\mathcal{F}(p^2)&\equiv&
1+\sum_{f=1}^{N_f}\frac{(m_0^2+\alpha p^2)/N_c}{p^2+M^2_{f f}}.
\end{eqnarray}
Note that for small sea-quark masses, $m_0$ can be taken infinity,
%%%%%%%%%%%%%% G(x) m_0 -> infty limit %%%%%%%%%%%%%%%%%%%%%%%
\begin{equation}
\bar{G}(M^2_{i i},M^2_{j j}|x) = \frac{1}{V}
\sum_{p\neq 0}
\frac{e^{ipx}}{(p^2+M^2_{ii})(p^2+M^2_{jj})
\left(\sum^{N_f}_f \frac{1}{p^2+M^2_{ff}}\right)} ~.
\end{equation}
Conversely,
in the quenched limit of taking the sea quark masses to infinity,
%$\to \infty$, 
$\bar{G}$ becomes
%%%%%%%%%%%%%%%% G(x) quenched limit %%%%%%%%%%%%%%%%%%%%%%%%%%
\begin{equation}
\bar{G}(M^2_{i i},M^2_{j j}|x) \equiv \frac{1}{V}\sum_{p\neq 0}
\frac{e^{ipx}(m_0^2+\alpha p^2)/N_c}{(p^2+M^2_{ii})(p^2+M^2_{jj})}.
\end{equation}

%%%%%%%%%%%%%%%%%%%%%%%%%%%%%%%%%%%%%%%%%%%%%%%%%%%%%%%%%%%
%%%%%%%%%%%%%%%%%%%%%%%%%%%%%%%%%%%%%%%%%%%%%%%%%%%%%%%%%%%
\section{The chiral condensate}
\label{sec:condensate}
\setcounter{equation}{0}
%%%%%%%%%%%%%%%%%%%%%%%%%%%%%%%%%%%%%%%%%%%%%%%%%%%%%%%%%%%
%%%%%%%%%%%%%%%%%%%%%%%%%%%%%%%%%%%%%%%%%%%%%%%%%%%%%%%%%%%

\begin{figure*}[tbhp]
  \centering
  \includegraphics[width=12cm]{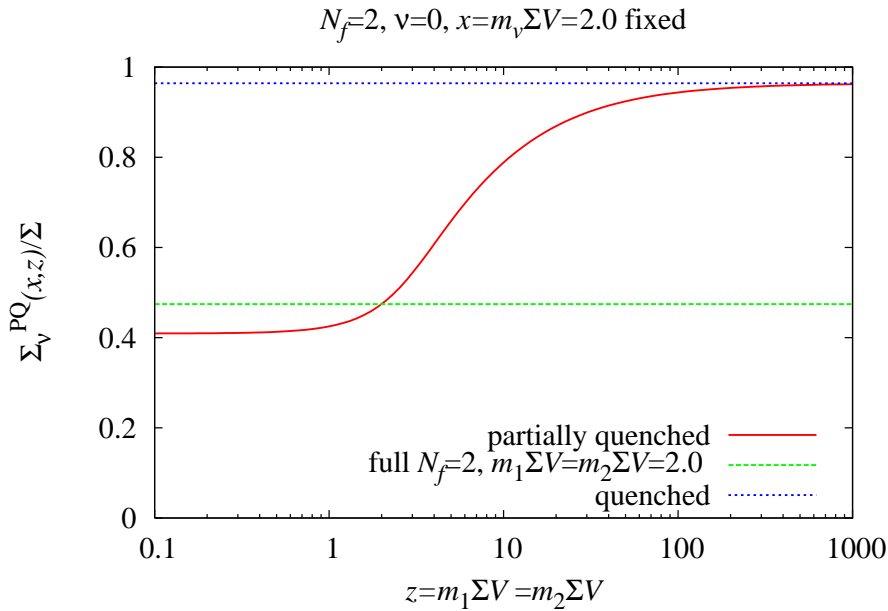}
  \caption{
    The sea quark mass dependence of the 2-flavor 
    partially quenched condensate at $\nu=0$ (solid). 
    The valence quark mass is fixed to $m_v\Sigma V=2.0$.
    As expected, we obtain a smooth curve crossing the full 2-flavor theory
    result (dashed) at $m_1=m_2=m_v$, and converging to the quenched limit (dotted).
  }
  \label{fig:PQcondensate}
\end{figure*}

At tree level, the partially 
quenched chiral condensate is obtained by the logarithmic
$x$-derivative of the
zero-mode partition function Eq.(\ref{eq:zero-mode}) followed
by the $y \to x$ limit,
%%%%%%%%%% condensate tree level %%%%%%%%%%%%%%%%%%%%%%%%%%
\begin{eqnarray}
\label{eq:sigma}
\frac{\Sigma^{{\rm PQ}}_\nu(x,\{z_i\})}{\Sigma}&=&\lim_{N\to 0} 
\frac{1}{N}\langle\sum_v^{N}[U_0+U_0^\dagger]_{v v}
\rangle_{U_0}
\nonumber\\
&=&
-\lim_{y\to x}\frac{\partial}{\partial x} 
\ln \mathcal{Z}^\nu_{1,1+N_f}(x | y, \{z_i\})
\nonumber\\
&=&
\frac{-1}{\mathcal{Z}^\nu_{0,N_f}(\{z_i\})
\prod_{i=1}^{N_f}(z_i^2-x^2)
\prod_{k>j}^{N_f}(z_k^2-z_j^2)}
\nonumber\\
&&\hspace{-0.5in}
\times \det \left(
\begin{array}{ccccc}
\partial_x K_\nu(x) & I_\nu(x) & I_\nu(z_1) & I_\nu(z_2) & \cdots\\
-\partial_x(x K_{\nu+1}(x)) & xI_{\nu+1}(x) & z_1I_{\nu+1}(z_1) & z_2I_{\nu+1}(z_2) & \cdots\\
\partial_x(x^2 K_{\nu+2}(x)) & x^2I_{\nu+2}(x) & z_1^2I_{\nu+2}(z_1) 
& z^2_2I_{\nu+2}(z_2) & \cdots\\
\cdots & \cdots & \cdots & \cdots & \cdots
\end{array}
\right),
\nonumber\\
\end{eqnarray}
where $x=m_v\Sigma V, z_i=m_i\Sigma V$,
and we use Eq.(\ref{eq:xylimit}) for the denominator and
the minus sign in the first line is due to the 
derivative with respect to the bosonic flavor. The quantity
$\Sigma^{{\rm PQ}}_\nu(x,\{z_i\})/\Sigma$ has two properties
that follow immediately from the corresponding statements
about the partition functions:
%%%%%%%%%%%% Properties of Sigma
\begin{enumerate}
%%%%%%%%%%%%%%%%%%%%%%%% item  N_f theory -> N_f-1 theory
\item 
When one of the sea quark masses is large, the partially
quenched chiral condensate reduces to 
%the partially quenched chiral condensate 
that of the $(N_f-1)$-flavor theory;
%%%%%%%%%%%%%% Eq N_f theory -> N_f-1 theory %%%%%%%%%%%%%
\begin{eqnarray}
\label{eq:condenque}
\Sigma^{{\rm PQ}}_\nu
(x,\{\underbrace{z_1,\cdots z_j\to \infty, \cdots, z_{N_f}}_{N_f \;\rm flavors}\})
/\Sigma=\hspace{1in} 
\nonumber\\
\Sigma^{{\rm PQ}}_\nu(x, 
\{\underbrace{z_1,\cdots,z_{j-1},z_{j+1}\cdots z_{N_f}}_{(N_f-1) \;\rm flavors}\})
/\Sigma.
\end{eqnarray}
As a particular case, one recovers the fully quenched condensate
\cite{Damgaard:2001js} 
when all sea quark masses are large, {\it i.e.} when $m_i \to \infty$
(See Fig.~\ref{fig:PQcondensate}).

%%%%%%%%%%%%%%%%%%%%%%%%% item full theory limit
\item 
If the valence quark mass $m_v$ is equal to one of sea quark masses $m_j$,
it is equivalent to the $j$-th flavor quark condensate in the full theory:
%%%%%%%%%%%%%%%% Eq full theory limit %%%%%%%%%%%%%%%%%%%
\begin{eqnarray}
\label{eq:condenfull}
\hspace{-0.5in}\lim_{x\to z_j}\frac{\Sigma^{{\rm PQ}}_\nu(x,\{z_i\})}{\Sigma}=
%\frac{\Sigma_\nu(\{z_i\})}{\Sigma}
\partial_{z_j} \ln Z^\nu_{N_f}(\{z_i\})
\equiv \frac{\Sigma_{\nu}^{{\rm full}\; (N_f, j)}(\{z_i\})}{\Sigma}
\;\;\;\mbox{(for any}\;\; 
j).\nonumber\\
\end{eqnarray}
\end{enumerate}

The first property follows directly from Eq.(\ref{eq:Qlim})
and the fact that
\begin{equation}
\mathcal{Z}^{\nu+N_f+1}_{0,1}/\mathcal{Z}^\nu_{0,1}\to 1
\end{equation}
in the large mass limit.
The second property can be shown explicitly by using 
Eq.(\ref{eq:xylimit}) and noting that
the product $\prod_{i=1}^{N_f}(z_i^2-x^2)\prod_{k>j}^{N_f}(z_k^2-z_j^2)$
is always antisymmetric under a swap of $x \leftrightarrow z_j$.
Together with the equation
%%%%%%%%%%%%%%%% Eq determinant swap %%%%%%%%%%%%%%%%%%%
\begin{eqnarray}
\det \left(
\begin{array}{ccccc}
K_\nu(x) & \partial_{z_j}I_\nu(z_j) & \cdots 
& \overbrace{I_\nu(z_j)}^{j\mbox{-th column}} & \cdots\\
-x K_{\nu+1}(x) & \partial_{z_j}(z_jI_{\nu+1}(z_j)) 
& \cdots & z_jI_{\nu+1}(z_j) & \cdots\\
x^2 K_{\nu+2}(x) & \partial_{z_j}(z_j^2I_{\nu+2}(z_j)) 
& \cdots
& z^2_jI_{\nu+2}(z_j) & \cdots\\
\cdots & \cdots & \cdots & \cdots & \cdots
\end{array}
\right)=\hspace{0.5in}
\nonumber\\
-\det \left(
\begin{array}{ccccc}
K_\nu(x) & I_\nu(z_j) & \cdots 
& \overbrace{\partial_{z_j}I_\nu(z_j)}^{j\mbox{-th column}} & \cdots\\
-x K_{\nu+1}(x) & z_jI_{\nu+1}(z_j) 
& \cdots & \partial_{z_j}(z_jI_{\nu+1}(z_j)) & \cdots\\
x^2 K_{\nu+2}(x) & z_j^2I_{\nu+2}(z_j)
& \cdots
& \partial_{z_j}(z^2_jI_{\nu+2}(z_j)) & \cdots\\
\cdots & \cdots & \cdots & \cdots & \cdots
\end{array}
\right),
\end{eqnarray}
which holds for any $j$, the statement follows.

For the normalization factor of the correlation functions to be
computed in the following sections we need to evaluate the
partition function to the given order in the expansion. This
essentially boils down to an evaluation of the one-loop
correction to the partially quenched chiral condensate. This
correction can be calculated in a standard manner:

%%%%%%%%%% Eq condensate 1-loop level %%%%%%%%%%%%%%%%
\begin{eqnarray}
\frac{\Sigma^{{\rm PQ}, \rm 1-loop}_\nu(x,\{z_i\})}{\Sigma}
&=&
\frac{1}{\Sigma V}
\lim_{N\to 0}\frac{1}{N}\frac{\partial}{\partial m_v} 
\ln Z^\nu_{N_f+N}(m_v,\{m_i\})
\nonumber\\
&&\hspace{-0.5in} 
=\lim_{N\to 0}\left(
1-\frac{1}{F^2NV}\sum^{N_f+N}_i\sum_v^{N}
\left\langle 
\int d^4 x 
%\bar{P}_{(vi)(iv)}(0)
\xi_{v i}(x)\xi_{i v}(x)\right\rangle_\xi
\right)
\nonumber\\
&&
\times \frac{1}{N}\langle\sum_v^{N}[U_0+U_0^\dagger]_{v v},
\rangle^{\rm 1-loop}_{U_0},
\end{eqnarray}
where $\langle \cdots \rangle_\xi$ denotes the integral over the
$\xi$ field, while the zero-mode integral 
$\langle \cdots\rangle^{\rm 1-loop}_{U_0}$ can be calculated 
as above but due to the vacuum bubble, the arguments  
$x$ and $z_i$ are shifted, at 1-loop, as
%%%%%%%%%%%%%% Eq x, z shifts %%%%%%%%%%%%%%%%%%%%%%%%%%%%%
\begin{eqnarray}
m_v \Sigma V =x \to x^{\rm eff}%m_v\Sigma^v_{\rm eff} V 
\;\;\;{\rm and}\;\;\;
m_i \Sigma V =z_i \to z^{\rm eff}_i, %m_i \Sigma^i_{\rm eff} V,
\end{eqnarray}
where
%%%%%%%%%%%%%% 1-loop corrections with xi %%%%%%%%%%%%%%%%%
\begin{eqnarray}
%\Sigma^v_{\rm eff}&=&\Sigma
\frac{x^{\rm eff}}{x} &=& 
\lim_{N\to 0}
\left(1-\frac{1}{F^2NV}\sum^{N_f+N}_i\sum_v^{N}
\langle \int d^4 x \xi_{i v}\xi_{vi}\rangle_\xi\right),
\nonumber\\
%\Sigma^i_{\rm eff}&=&\Sigma\times
\frac{z^{\rm eff}_i}{z_i}&=&
\lim_{N\to 0}
\left(
1-\frac{1}{F^2V}\sum^{N_f+N}_f
\langle \int d^4 x \xi_{i f}\xi_{fi}\rangle_\xi\right).
\end{eqnarray}

With the Feynmann rules given in the previous section,
the $\xi$ integral can be performed, and we obtain
\cite{Osborn:1998qb}
%%%%%%%%%%%% Eq 1-loop correction general %%%%%%%%%%%%%%%
\begin{eqnarray}
%\Sigma^v_{\rm eff}&=&\Sigma-\frac{\Sigma}{F^2}
\frac{x^{\rm eff}}{x} &=& 1-\frac{1}{F^2}
\left(
\sum_i^{N_f}\bar{\Delta}(M^2_{i v}|0)
-\bar{G}(M^2_{v v},M^2_{v v}|0)
%-\frac{1}{V}\sum_{p\neq 0} 
%\frac{(m_0^2+\alpha p^2)/N_c}{(p^2+M_{v v}^2)^2\mathcal{F}(p^2)}
\right),
\nonumber\\
%\Sigma^i_{\rm eff}&=&\Sigma-\frac{\Sigma}{F^2}
\frac{z^{\rm eff}_i}{z_i}&=&1-\frac{1}{F^2}
\left(
\sum_{j}^{N_f}\bar{\Delta}(M^2_{i j}|0)
-\bar{G}(M^2_{i i},M^2_{i i}|0)
%\frac{1}{V}\sum_{p \neq 0}
%\frac{(m_0^2+\alpha p^2)/N_c}{(p^2+M_{i i}^2)^2\mathcal{F}(p^2)}
\right) ~.
\end{eqnarray}
The corrections are UV divergent,
and thus need regularization.
Note that each of $x$ and $z_i$ receive different one-loop 
correction in general.
In practice, however, the following three special cases 
are of our interest:
\begin{enumerate}
\item Both of the valence quarks and the sea quarks 
are in the $\epsilon$-regime:\\
One can to this order set $M_{i j}=0$ for all $i$ and $j$, and 
take the $m_0\to \infty$ limit, which leads to
%%%%%%%%%%%%% Eq 1-loop correction e-regime %%%%%%%%%%%%%
\begin{equation}
\label{eq:sigmaeff-full-deg}
%\Sigma^v_{\rm eff}/\Sigma=\Sigma^i_{\rm eff}/\Sigma=
\frac{x^{\rm eff}}{x}=\frac{z^{\rm eff}_i}{z_i}=
1-\frac{1}{F^2}\frac{N^2_f-1}{N_f}\bar{\Delta}(0|0)\;\;
\mbox{(for all $i$)} ~.
\end{equation}
%As expected, 
This correction is equivalent to that of $N_f$-flavor full theory.
In dimensional regularization,
%%%%%%%%%%%% Eq beta_1 %%%%%%%%%%%%%%%%%%%%%%%%%%%%%%%%%%
\begin{equation}
\bar{\Delta}(0|0) = -\frac{\beta_1}{L^2}+{\cal O}(1/L^4),
\end{equation}
is obtained where $\beta_1$ is known as the shape coefficient 
\cite{Hasenfratz:1989pk}. It
depends only on the shape of the 4-dimensional 
Euclidean space-time volume.
\item The sea quarks are in the $p$-regime while 
the valence quarks are still in the $\epsilon$-regime
(the $m_i$'s are heavy but much smaller than the QCD scale, 
$\Lambda_{\rm QCD} )$:\\
One can take the $m_0\to \infty $ limit but should keep 
%$m_v$ %{\bf why should $m_v$ be kept ``finite''??}
%and 
$m_i$'s finite, which leads to
%%%%%%%%%%%%%%%% Eq 1-loop correction p-regime %%%%%%%%%%
\begin{eqnarray}
%\Sigma^v_{\rm eff}/\Sigma
\frac{x^{\rm eff}}{x}&=&
1-\frac{1}{F^2}\left(
\sum_i^{N_f}\bar{\Delta}(M^2_{i v}|0)
-\frac{1}{V}\sum_{p\neq 0}
\frac{1}{p^4%(p^2+M_{v v}^2)^2
(\sum^{N_f}_i\frac{1}{p^2+M^2_{i i}})}\right),
\nonumber\\
%\Sigma^i_{\rm eff}/\Sigma
\frac{z^{\rm eff}_i}{z_i}
&=&
1-\frac{1}{F^2}\left(
\sum_j^{N_f}\bar{\Delta}(M^2_{i j}|0)
-\frac{1}{V}\sum_{p\neq 0}
\frac{1}{(p^2+M_{i i}^2)^2(\sum^{N_f}_j\frac{1}{p^2+M^2_{j j}})}\right).
\nonumber\\
\end{eqnarray}
Note that a double pole contribution appears in $x^{\rm eff}/x$, 
as an effect of the partially quenching.
% if $m_v=0$ 
%with all $m_i$'s being finite.
%unless the $m_v$ and $m_i$'s are all degenerate.
\item All the sea quark masses are heavy, $m_i\gg \Lambda_{\rm QCD}$:\\
In this case, $m_0$ cannot be large, but one can take the 
$m_i\to \infty$ limit, which leads to
%%%%%%%%%%% Eq 1-loop correction quenched %%%%%%%%%%%%%%%
\begin{equation}
%\Sigma^v_{\rm eff}/\Sigma=
\frac{x^{\rm eff}}{x}=
1+\frac{1}{F^2}\left(
\frac{1}{V}\sum_{p\neq 0}
\frac{(m_0^2+\alpha p^2)/N_c}{p^4}\right),
\end{equation}
which agrees with the quenched result.
\end{enumerate}

To summarize this section, the chiral condensate to one-loop order is 
given by
%%%%%%%%%%%%% Eq summary of condensate %%%%%%%%%%%%%%%%%%
\begin{equation}
\Sigma^{{\rm PQ}, \rm 1-loop}_\nu(x,\{z_i\})
= 
\Sigma^{{\rm PQ}}_\nu(x^{\rm eff},\{z^{\rm eff}_i\})\frac{x^{\rm eff}}{x},
\end{equation}
where the analytical functional form of 
$\Sigma^{{\rm PQ}}_\nu(x, \{z_i\})$ is given by Eq.(\ref{eq:sigma}).
When all the quarks are in the $\epsilon$-regime, the one-loop correction
is, to this order, constant, and can simply be taken into account in
the Lagrangian by shifting $\Sigma$ according to the above prescription. 
The chiral condensate in the infinite volume limit, $\Sigma$,
and all the other low-energy constants, are of course expected to
depend on the number of flavors.
Matching conditions 
\cite{Giusti:2004an, Hernandez:2006kz, DeGrand:2006uy, 
DeGrand:2006nv, Hasenfratz:2007yj}
can ensure
smooth connections between theories with different number of
flavors.

%%%%%%%%%%%%%%%%%%%%%%%%%%%%%%%%%%%%%%%%%%%%%%%%%%%%%%%%%%%%%%%%%%%%
%%%%%%%%%%%%%%%%%%%%%%%%%%%%%%%%%%%%%%%%%%%%%%%%%%%%%%%%%%%%%%%%%%%%
\section{Zero-mode integrals in the partially quenched theory}
\label{sec:zero-modes}
\setcounter{equation}{0}
%%%%%%%%%%%%%%%%%%%%%%%%%%%%%%%%%%%%%%%%%%%%%%%%%%%%%%%%%%%%%%%%%%%%
%%%%%%%%%%%%%%%%%%%%%%%%%%%%%%%%%%%%%%%%%%%%%%%%%%%%%%%%%%%%%%%%%%%%

%%%%%%%%%%%%%%%%%%%%%%%%%%%%%%%%%%%%%%%%%%%%%%%%%%%%%%%%%%%%
\subsection{$U(N_f+N)$ group integrals in the replica limit}
\label{sec:Uint}
%%%%%%%%%%%%%%%%%%%%%%%%%%%%%%%%%%%%%%%%%%%%%%%%%%%%%%%%%%%%

In the $\epsilon$-regime, the integral over the zero-mode, $U_0$
(which for simplicity of notation will be denoted by $U$
in this subsection) has to be done 
exactly. This is the central difference between the 
$\epsilon$-regime and the $p$-regime, and we are fortunately able
to perform the required group integrations exactly. Because
these technical aspects are so important for the calculation
that follows, we first give a detailed outline of how the
integrations have been done, and how we employ the replica formalism
in this context. 
%In this subsection we calculate the $U(N_f+N)$ group integrals 
%in the replica limit $N\to 0$ for various matrix elements,
%which is an essential part in order to treat non-degenerate 
%sea and valence quarks.
As in Ref. \cite{Damgaard:2001js} the needed group integrals
are conveniently obtained through the identities that follow
from the fact that  
$[\det (\Sigma V\mathcal{M})]^{-\nu}
\mathcal{Z}^\nu_{N_f+N}(x,\{z_i\})$ is a function of 
$\mathcal{M}^\dagger\mathcal{M}$ only. Since the group
integrals are known for diagonal sources, this basically solves
the problem.
We begin, however, with two simpler cases. They 
do not require any special techniques, and can be evaluated
straightforwardly by making use of the graded partition function:
\begin{eqnarray}
%\lim_{N\to 0}\frac{1}{N}\sum_v
%%%%%%%%%%%%%%% Eq (U+U) condensate %%%%%%%%%%%%%%%%%%
\frac{1}{2}\langle (U_{v v}+U^\dagger_{v v}) \rangle
&=&\lim_{y\to x}
\partial_y \ln \mathcal{Z}^\nu_{1,1+N_f}(x|y,\{z_i\})=
\frac{\Sigma^{{\rm PQ}}_{\nu}(x,\{z_i\})}{\Sigma},\\
%\lim_{N\to 0}\frac{1}{N}\sum_v
%%%%%%%%%%%%%%% Eq (U+U)^2 %%%%%%%%%%%%%%%%%%%%%%%%%%%%%
\frac{1}{4}\langle 
(U_{v v}+U_{v v}^\dagger)^2 \rangle
&=&\frac{1}{\mathcal{Z}^\nu_{N_f}(\{z_i\})}
\lim_{y\to x}\partial^2_y \mathcal{Z}^\nu_{1,1+N_f}(x|y,\{z_i\})
\nonumber\\
&=&
\frac{\partial_x\Sigma^{{\rm PQ}}_{\nu}(x,\{z_i\})}{\Sigma}
-\frac{\Delta \Sigma^{{\rm PQ}}_{\nu}(x,\{z_i\})}{\Sigma},
\label{eq:diag+}
\end{eqnarray}
where the second term of Eq.(\ref{eq:diag+}) is defined by
%%%%%%%%%%%% Eq Delta Sigma definition %%%%%%%%%%%%%%%%%%%
\begin{eqnarray}
\frac{\Delta \Sigma^{{\rm PQ}}_{\nu}(x,\{z_i\})}{\Sigma}
\equiv \frac{\lim_{y\to x}
\partial_y \partial_x 
\mathcal{Z}^\nu_{1,1+N_f}(x|y,\{z_i\})}
{\mathcal{Z}^\nu_{N_f}(\{z_i\})}.
\end{eqnarray}
If we put $x = z_j$ (for any $j$), this simply amounts to
removing the fermion determinant of quark species $j$. From
the definition (\ref{eq:diag+}) we then immediately get
%%%%%%%%%%%%% Eq (U+U)^2 full theory limit %%%%%%%%%%%%%%%
\begin{eqnarray}\left.\left[
\frac{\partial_x\Sigma^{{\rm PQ}}_{\nu}(z_j,\{z_i\})}{\Sigma}
-\frac{\Delta \Sigma^{{\rm PQ}}_{\nu}(z_j,\{z_i\})}{\Sigma}
\right.\right]_{x=z_j}
%{\to \atop {x\to z_j}}\right]\right|_{x = z_j} 
%\xrightarrow[x \to z_j]{}
%\stackrel{x \to z_j}{\longrightarrow}
=
\hspace{1.5in}\nonumber\\ 
\partial_{z_j} 
\frac{\Sigma^{{\rm full}\; (N_f, j)}_{\nu}(\{z_i\})}{\Sigma}
+\left(
\frac{\Sigma^{{\rm full}\;(N_f, j)}_{\nu}(\{z_i\})}{\Sigma}
\right)^2 ~,
\end{eqnarray}
corresponding to the result in the full theory without partial
quenching.
%which is consistent with the full theory.
%Here we have used 
%%%%%%%%%%%%% Eq PQ Sigma and full Sigma %%%%%%%%%%%%%%%%%%
%\[
%\frac{\Sigma^{{\rm PQ}}_{\nu}(z_j+\epsilon,\{\cdots,z_j+\epsilon,\cdots\})}
%{\Sigma}
%=\frac{\Sigma^{{\rm full}\;(N_f, j)}_{\nu}(\{\cdots,z_j+\epsilon,\cdots\})}{\Sigma}.
%\]
Conversely, in
the limit $z_i\to \infty$ (for all $i$), 
%%%%%%%%%%%%% Eq (U+U)^2 quenched limit %%%%%%%%%%%%%%%%%%%
\begin{eqnarray}
\frac{\partial_x\Sigma^{{\rm PQ}}_{\nu}(x,\{z_i\})}{\Sigma}
&
\xrightarrow[{\{z_i\}\to \infty}]{}
& 
\frac{\partial_x\Sigma^{que}_{\nu}(x)}{\Sigma}, \nonumber\\
-\frac{\Delta \Sigma^{{\rm PQ}}_{\nu}(x,\{z_i\})}{\Sigma}
&\xrightarrow[{\{z_i\}\to \infty}]{} & 
1+\frac{\nu^2}{x^2},
\label{eq:DeltaSigmaquenchlimit}
\end{eqnarray}
we recover the results of the quenched theory \cite{Damgaard:2001js}.
Here, in a hopefully obvious notation, we have
denoted the chiral condensates in the full and quenched theories
by $\Sigma_\nu^{{\rm full}\; (N_f, j)}$ and $\Sigma_\nu^{que}$ 
(See Eq.(\ref{eq:condenque}) and Eq.(\ref{eq:condenfull})),
respectively.\\

We
next consider a purely imaginary source $iJ$ (with $J$ real)  
on a diagonal $(v,v)$ element of $\mathcal{M}$,
%%%%%%%%%%%%%% Eq (v,v) imag source %%%%%%%%%%%%%%%%
\begin{equation}
\label{eq:srci1}
\Sigma V \mathcal{M+J}
={\rm diag}(\underbrace{z_1,z_2,\cdots}_{N_f}, 
\underbrace{(x+iJ),x,\cdots, x}_{N}).
\end{equation}
We note that 
%%%%%%%%%%%%%% Eq mass matrix determinant %%%%%%%%%%%
\begin{equation}
\lim_{N \to 0}\det (\Sigma V\mathcal{M+J})= 
\det (\Sigma V \mathcal{M})\times(1+iJ/x),
\end{equation}
%$(\Sigma V\mathcal{M+J})^\dagger (\Sigma V\mathcal{M+J})$ 
and
%%%%%%%%%%%%%% Eq M^2 diagonalization %%%%%%%%%%%%%%%
\begin{equation}
(\Sigma V\mathcal{M+J})^\dagger (\Sigma V\mathcal{M+J})
= {\rm diag}(\underbrace{z^2_1,z^2_2,\cdots}_{N_f}, 
\underbrace{\lambda^2,x^2,\cdots, x^2}_{N}),
\end{equation}
where $\lambda=\sqrt{x^2+J^2}$.  By the chain rule,  
$J$-derivative and $\lambda$-derivative are related through
%%%%%%%%%%%%% Eq J-derivative and lambda-derivative %%%%%
\begin{equation}
\left.\frac{\partial}{\partial J}\right|_{J=0}
=\left.\frac{J}{\lambda}\frac{\partial}{\partial \lambda}\right|_{J=0}=0,\;\;\;
\left.\frac{\partial^2}{\partial J^2}\right|_{J=0}
=\left.\frac{1}{x}\frac{\partial}{\partial \lambda}\right|_{\lambda=x}.
\end{equation}
From the above equations, we obtain
(denote $\mathcal{Z}^\nu_{N_f+N}(x,\{z_i\})$ by $\mathcal{Z}$ for simplicity),
%%%%%%%%%%%%% Eq J and lambda equations %%%%%%%%%%%%% 
\begin{eqnarray}
\hspace{-0.1in}\frac{1}{\mathcal{Z}}
\frac{\partial}{\partial J}
\det (\Sigma V\mathcal{M+J})^{-\nu}
\mathcal{Z}|_{J=0}
&=&\det (\Sigma V\mathcal{M})^{-\nu} 
\left(-\frac{i\nu}{x}+\frac{1}{\mathcal{Z}}
\frac{\partial}{\partial J}\mathcal{Z}\right)=0,
\nonumber\\\hspace{-0.5in}
\frac{1}{\mathcal{Z}}
\frac{\partial^2}{\partial J^2}
\det (\Sigma V\mathcal{M+J})^{-\nu}
\mathcal{Z}|_{J=0}
&=&\det (\Sigma V\mathcal{M})^{-\nu}\left(
\frac{\nu(\nu-1)}{x^2}
+\left.\frac{1}{\mathcal{Z}}\frac{\partial^2 \mathcal{Z}}
{\partial J^2}\right|_{J=0}
\right),
\nonumber\\\hspace{-0.5in}
\frac{1}{\mathcal{Z}}\frac{1}{x}\frac{\partial}{\partial \lambda}
\det (\Sigma V\mathcal{M+J})^{-\nu}
\mathcal{Z}|_{\lambda=x}
&=&\det (\Sigma V\mathcal{M})^{-\nu}
\left(-\frac{\nu}{x^2}+\frac{\Sigma^{{\rm PQ}}_\nu(x,\{z_i\})}{x\Sigma}\right) ~.
\nonumber\\
\end{eqnarray}
This gives us the useful identities
\begin{eqnarray}
\label{eq:axial}
%%%%%%%%%%%%% Eq (U-U) %%%%%%%%%%%%%%%%%%%%%%%%%%%%
%\lim_{N\to 0}\frac{1}{N}\sum_v
\frac{1}{2}\langle (U_{v v}-U^\dagger_{v v}) \rangle
&=& \left.i\frac{1}{\mathcal{Z}}\frac{\partial \mathcal{Z}}
{\partial J}\right|_{J=0}=-\frac{\nu}{x}\\
\label{eq:diag-}
%%%%%%%%%%%%% Eq (U-U)^2 %%%%%%%%%%%%%%%%%%%%%%%%%%
%\lim_{N\to 0}\frac{1}{N}\sum_v
\frac{1}{4}\langle 
(U_{v v}-U_{v v}^\dagger)^2 \rangle
&=& -\left.\frac{1}{\mathcal{Z}}\frac{\partial^2 \mathcal{Z}}
{\partial J^2}\right|_{J=0}
= -\frac{\Sigma^{{\rm PQ}}_\nu(x,\{z_i\})}{x\Sigma} + \frac{\nu^2}{x^2} ~.
\end{eqnarray} 
%The source $\mathcal{J}$ in Eq.(\ref{eq:srci1})
%also leads to an equation
%%%%%%%%%%%%%% Eq (U^2-U^2) %%%%%%%%%%%%%%%%%%%%%%%%%
%\begin{eqnarray}
%%\lim_{N\to 0}\frac{1}{N}\sum_v
%\frac{1}{4}\langle 
%(U_{v v})^2-(U_{v v}^\dagger)^2 \rangle
%&=&\frac{1}{\mathcal{Z}}\lim_{N\to 0}\frac{1}{N}
%\frac{\partial}{\partial x}
%\left.\left[\frac{i\partial}{\partial J}\mathcal{Z}\right]\right|_{J=0}
%\nonumber\\
%&=&\frac{1}{\mathcal{Z}}\lim_{y \to x}
%\frac{\partial}{\partial y}
%\left[-\frac{\nu}{y}\mathcal{Z}^\nu_{1,1+N_f}(x,y,\{z_i\})\right]
%\nonumber\\
%&=&\frac{\nu}{x^2}-\frac{\nu\Sigma^{{\rm PQ}}_\nu(x,\{z_i\})}{x\Sigma}.
%\end{eqnarray}

In order to calculate the meson correlators in PQChPT, we also need
matrix elements which have different valence flavor indices. For
example,
for the disconnected correlators, we need
%%%%%%%%%%% (U+U)(U+U) %%%%%%%%%%%%%%%%%%%%%%%%%%%%%%%%
\begin{eqnarray}
\label{eq:DPQ}
\frac{1}{4}\langle 
(U_{v_1 v_1}+U_{v_1 v_1}^\dagger) (U_{v_2 v_2}+U_{v_2 v_2}^\dagger) \rangle
%\hspace{0.5in}
\nonumber\\
&&
\hspace{-2in}=\frac{1}{\mathcal{Z}^\nu_{N_f}(\{z_i\})}
\lim_{y_1\to x_1, y_2\to x_2}
\partial_{y_1}\partial_{y_2} 
\mathcal{Z}^\nu_{2,2+N_f}(x_1,x_2|y_1,y_2,\{z_i\})
\nonumber\\
&&\hspace{-2in}\equiv D_\nu^{{\rm PQ}}(x_1,x_2,\{z_i\}).
%\nonumber\\
\end{eqnarray}

To derive the analytical expressions for these and others closely
related, we first consider two purely imaginary sources
$\mathcal{J}_{v_1 v_1}=iJ_1$  
and $\mathcal{J}_{v_2 v_2}=iJ_2$ along the diagonal: 
%%%%%%%%%%%% Eq sources (1 1) iJ_1 (2 2) iJ_2 %%%%%%%%%
\begin{equation}
\Sigma V \mathcal{M+J}
={\rm diag}(\underbrace{z_1,z_2,\cdots}_{N_f}, 
\underbrace{(x_1+iJ_1),(x_2+iJ_2),x,\cdots, x}_{N}),
\end{equation}
The replica limit of this determinant is
%%%%%%%%%%%% Eq determinant %%%%%%%%%%%%%%%%%%%%%%%%%%%
\begin{equation}
\lim_{N \to 0}\det (\Sigma V\mathcal{M+J})= 
\det (\Sigma V \mathcal{M})\times(1+iJ_1/x_1)(1+iJ_2/x_2),
\end{equation}
and $(\Sigma V\mathcal{M+J})^\dagger (\Sigma V\mathcal{M+J})$ 
takes the form
%%%%%%%%%%%% Eq diagonalization %%%%%%%%%%%%%%%%%%%%%%
\begin{equation}
(\Sigma V\mathcal{M+J})^\dagger (\Sigma V\mathcal{M+J})
\to {\rm diag}(\underbrace{z^2_1,z^2_2,\cdots}_{N_f}, 
\underbrace{\lambda_1^2,\lambda_2^2,x^2,\cdots, x^2}_{N}),
\end{equation}
where $\lambda_i=\sqrt{x_i^2+J_i^2}(i=1,2)$. Again,  
$J$-derivative and $\lambda$-derivative are related through
the chain rule,
%%%%%%%%%%%% Eq J-deriv and lambda deriv %%%%%%%%%%%%%%%
\begin{equation}
\left.\frac{\partial \lambda_i}{\partial J_i}\right|_{J_i=0}
=\left.\frac{J_i}{\lambda_i}\right|_{J_i=0}=0.
\end{equation}
{}From these equations we get 
%%%%%%%%%%%% Eq (U-U)(U-U) %%%%%%%%%%%%%%%%%%%%%%%%%%%%%
\begin{eqnarray}
\left.\frac{1}{\mathcal{Z}}
\frac{\partial^2}{\partial J_1 \partial J_2}
\det (\Sigma V\mathcal{M+J})^{-\nu}
\mathcal{Z}\right|_{\mathcal{J}=0}\hspace{2in}\nonumber\\
=\det (\Sigma V\mathcal{M})^{-\nu}\left(
\frac{-\nu^2}{x_1x_2}
+\left.\frac{1}{\mathcal{Z}}\frac{\partial^2 \mathcal{Z}}
{\partial J_1 \partial J_2}\right|_{\mathcal{J}=0}
\right)=0,\nonumber\\
%\lim_{N\to 0}\frac{1}{N}\sum_v
\frac{1}{4}\langle 
(U_{v_1 v_1}-U_{v_1 v_1}^\dagger) (U_{v_2 v_2}-U_{v_2 v_2}^\dagger)\rangle
= \frac{\nu^2}{x_1x_2} ~.
\end{eqnarray}

Next let us consider a real off-diagonal source,
$\mathcal{J}_{v_1 v_2}=J$, for which  
the determinant is unchanged in the replica limit:
%%%%%%%%%% Eq source (1 2) J %%%%%%%%%%%%%%%%%%%%%%%%%
\begin{equation}
\lim_{N\to 0}\det (\Sigma V\mathcal{M+J})= 
\det (\Sigma V \mathcal{M}) ~.
\end{equation}
Now $(\Sigma V\mathcal{M+J})^\dagger (\Sigma V\mathcal{M+J})$ 
can be diagonalized as
%%%%%%%%%% Eq diagonalization %%%%%%%%%%%%%%%%%%%%%%%%
\begin{equation}
(\Sigma V\mathcal{M+J})^\dagger (\Sigma V\mathcal{M+J})
\to {\rm diag}(\underbrace{z^2_1,z^2_2,\cdots}_{N_f}, 
\underbrace{\lambda_+^2,\lambda_-^2,x^2,\cdots, x^2}_{N}) ~,
\end{equation}
where
%%%%%%%%%% Eq lambda_\pm %%%%%%%%%%%%%%%%%%%%%%%%%%%%%
\begin{eqnarray}
\lambda_\pm=
\sqrt{\frac{(J^2+x_1^2+x_2^2)\pm 
\sqrt{J^4+2J^2(x_1^2+x_2^2)+(x_1^2-x_2^2)^2}}{2}}.
\end{eqnarray}
If we assume that $x_1 \neq x_2$ (the special case $x_1 \to x_2$
\cite{Damgaard:2001js} can be taken as a limiting case afterwards,
see below), we obtain the following relation between 
$J$-derivatives and $\lambda$-derivatives: 
%%%%%%%%%% Eq J - lambda derivatives %%%%%%%%%%%%%%%%%
\begin{equation}
\left.\frac{\partial^2 }{\partial J^2}\right|_{J=0}
=\frac{1}{x_1^2-x_2^2}
\left.\left(x_1\frac{\partial}{\partial \lambda_+}\right|_{\lambda_+=x_1}
-x_2\left.\frac{\partial}{\partial \lambda_-}\right|_{\lambda_-=x_2}
\right),
\label{eq:J-lambda}
\end{equation}
We now use
%%%%%%%%%% Eq lambda-derivative %%%%%%%%%%%%%%%%%%%%%%
\begin{eqnarray}
\left.\frac{1}{\mathcal{Z}}
\frac{1}{x_1^2-x_2^2}
\left(x_1\frac{\partial}{\partial \lambda_+}
-x_2\frac{\partial}{\partial \lambda_-}
\right)
\mathcal{Z}\right|_{\lambda_+=x_1,\lambda_-=x_2}
=\hspace{1in}\nonumber\\
\frac{1}{x_1^2-x_2^2}\left(x_1\frac{\Sigma_{\nu}^{{\rm PQ}}(x_1,\{z_i\})}{\Sigma}
-x_2\frac{\Sigma_{\nu}^{{\rm PQ}}(x_2,\{z_i\})}{\Sigma}\right) ~.
\end{eqnarray}
Note that a purely imaginary source $\mathcal{J}_{v_1 v_2}=iJ$ gives
the same results. Thus, one obtains
%%%%%%%%%% Eq (U+U)^2 2-valence %%%%%%%%%%%%%%%%%%%%%%
\begin{equation}
%\hspace{-0.2in}
\frac{1}{4}\langle 
(U_{v_1 v_2}\pm U_{v_2 v_1}^\dagger)^2\rangle
= 
\frac{\pm}{x_1^2-x_2^2}
\left(x_1\frac{\Sigma_{\nu}^{{\rm PQ}}(x_1,\{z_i\})}{\Sigma}
-x_2\frac{\Sigma_{\nu}^{{\rm PQ}}(x_2,\{z_i\})}{\Sigma}\right).
\end{equation}
As a check, if we take the mass-degenerate limit $x_2 \to x_1$ we
recover 
\begin{equation}
%\hspace{-0.2in}
\frac{1}{4}\langle 
(U_{v_1 v_2}\pm U_{v_2 v_1}^\dagger)^2\rangle
= 
\pm\frac{1}{2}
\left(\frac{\Sigma_{\nu}^{{\rm PQ}}(x_1,\{z_i\})}{x_1\Sigma}
+\frac{\partial_{x_1}\Sigma_{\nu}^{{\rm PQ}}(x_1,\{z_i\})}{\Sigma}\right),
\end{equation}
which is obtained from the formula in the degenerate case in
Eq.(\ref{eq:J-lambda}), 
\begin{equation}
\left.\frac{\partial^2 }{\partial J^2}\right|_{J=0}
=\left.\frac{1}{4x_1}\left(\frac{\partial}{\partial \lambda_+}
+\frac{\partial}{\partial \lambda_-}\right)
+\frac{1}{4}\left(
\frac{\partial}{\partial \lambda_+}
-\frac{\partial}{\partial \lambda_-}\right)^2
\right|_{\lambda_{\pm}=x_1},
\end{equation}
where $\lambda_{\pm}= (\sqrt{J^2+4x_1^2}\pm J)/2$. 

We finally put two real sources on the off-diagonal elements,
$\mathcal{J}_{v_1 v_2}=J_1$ and $\mathcal{J}_{v_2 v_1}=J_2$.
In the replica limit the determinant becomes
%%%%%%%%%% Eq sources (1 2) J_1 (2 1) J_2 %%%%%%%%%%%
\begin{equation}
\lim_{N\to 0}\det (\Sigma V\mathcal{M+J})= 
\det (\Sigma V \mathcal{M})(1-J_1J_2/x_1x_2),
\end{equation}
and $(\Sigma V\mathcal{M+J})^\dagger (\Sigma V\mathcal{M+J})$ 
diagonalizes as
%%%%%%%%%% Eq diagonalization %%%%%%%%%%%%%%%%%%%%%%%
\begin{equation}
(\Sigma V\mathcal{M+J})^\dagger (\Sigma V\mathcal{M+J})
\to {\rm diag}(\underbrace{z^2_1,z^2_2,\cdots}_{N_f}, 
\underbrace{\lambda_+^2,\lambda_-^2,x^2,\cdots, x^2}_{N}),
\end{equation}
where 
%%%%%%%%%% Eq lambda_\pm %%%%%%%%%%%%%%%%%%%%%%%%%%%%
\begin{eqnarray}
\lambda_\pm=
\sqrt{\frac{(J_1^2+J_2^2+x_1^2+x_2^2)\pm 
\sqrt{(J_1^2+J_2^2+x_1^2+x_2^2)^2-4(J_1J_2-x_1x_2)^2}}{2}}.
\nonumber\\
\end{eqnarray}
The relation between $J$-derivative and $\lambda$-derivative 
can be worked out as above. Assuming again that $x_1 \neq x_2$
(the degenerate case can also here be recovered by taking the
limit $x_1 \to x_2$ afterwards), we get 
%%%%%%%%%% Eq J-lambda derivatives %%%%%%%%%%%%%%%%%%
\begin{equation}
\left.\frac{\partial^2 }{\partial J_1\partial J_2}\right|_{J_i=0}
=\frac{1}{x_1^2-x_2^2}
\left.\left(x_2\frac{\partial}{\partial \lambda_+}\right|_{\lambda_+=x_1}
-x_1\left.\frac{\partial}{\partial \lambda_-}\right|_{\lambda_-=x_2}
\right).
\end{equation}
In the same way as above, we thus find that the two equations 
%%%%%%%%%% Eq J and lambda equations %%%%%%%%%%%%%%%%
\begin{eqnarray}
\left.\frac{1}{\mathcal{Z}}
\frac{\partial^2}{\partial J_1 \partial J_2}
\det (\Sigma V\mathcal{M+J})^{-\nu}
\mathcal{Z}\right|_{\mathcal{J}=0}
=\hspace{2in}
\nonumber\\
\det (\Sigma V\mathcal{M})^{-\nu}\left(
\frac{\nu}{x_1x_2}
+\frac{1}{\mathcal{Z}}\left.\frac{\partial^2 \mathcal{Z}}
{\partial J_1 \partial J_2}\right|_{\mathcal{J}=0}
\right),
\nonumber\\
%\lim_{N\to 0}\frac{1}{N}\sum_v
\frac{1}{\mathcal{Z}}
\frac{1}{x_1^2-x_2^2}
\left(x_2\frac{\partial}{\partial \lambda_+}
-x_1\frac{\partial}{\partial \lambda_-}
\right)
\det (\Sigma V\mathcal{M+J})^{-\nu}
\mathcal{Z}|_{\lambda_i=x_i}
=\det (\Sigma V\mathcal{M})^{-\nu}
\nonumber\\
\times\left(
\frac{\nu}{x_1x_2}
+\frac{1}{x_1^2-x_2^2}
\left(
(x_2\frac{\Sigma_{\nu}^{{\rm PQ}}(x_1,\{z_i\})}{\Sigma}
-x_1\frac{\Sigma_{\nu}^{{\rm PQ}}(x_2,\{z_i\})}{\Sigma}
\right)
\right),
\nonumber\\
\end{eqnarray}
lead to
%%%%%%%%% Eq (U+U)(U+U) 2valences %%%%%%%%
\begin{eqnarray}
\frac{1}{4}\langle 
(U_{v_1 v_2}\pm U_{v_2 v_1}^\dagger)(U_{v_2 v_1}\pm U_{v_1 v_2}^\dagger)\rangle
= \hspace{1in}\nonumber\\
\frac{1}{x_1^2-x_2^2}\left(x_2\frac{\Sigma_{\nu}^{{\rm PQ}}(x_1,\{z_i\})}{\Sigma}
-x_1\frac{\Sigma_{\nu}^{{\rm PQ}}(x_2,\{z_i\})}{\Sigma}\right),
\end{eqnarray}
where we used that purely imaginary sources
$\mathcal{J}_{v_1 v_2}=iJ_1$ $\mathcal{J}_{v_2 v_1}=-iJ_2$ give
the same result.
We summarize all pertinent formulas in appendix \ref{app:Uint}.\\

%%%%%%%%%%%%%%%%%%%%%%%%%%%%%%%%%%%%%%%%%%%%%%%%%%
\subsection{The Unitarity Formula}
%%%%%%%%%%%%%%%%%%%%%%%%%%%%%%%%%%%%%%%%%%%%%%%%%%

{}From the above formulae and the requirement of unitarity one finds 
in the replica limit,
%%%%%%%%%%% Eq 1=UU %%%%%%%%%%%%%%%%%%%%%%%%%%
\begin{eqnarray}
1&=&\left\langle\lim_{N\to 0} \sum^{N+N_f}_{i}U^\dagger_{v_1i}U_{i v_1}\right\rangle\nonumber\\
&=&\left\langle\lim_{N\to 0}(U^\dagger_{v_1 v_1}U_{v_1 v_1}+
\underbrace{U^\dagger_{v_1 v_2}U_{v_2 v_1}
+\cdots +U^\dagger_{v_1 v_{N}}U_{v_{N} v_1}}_{N-1})
+\sum^{N_f}_iU^\dagger_{v_1 i}U_{i v_1}\right\rangle
\nonumber\\
&=&\left\langle U^\dagger_{v_1 v_1}U_{v_1 v_1} - U^\dagger_{v_1 v_2}U_{v_2 v_1}
+\sum^{N_f}_iU^\dagger_{v_1 i}U_{i v_1}\right\rangle,
\end{eqnarray}
where we have used that the $N$ replicated quarks are all degenerate:
$x_i=m_{v_i}\Sigma V=x$.
One then obtains a non-trivial identity,
%%%%%%%%%%% Eq unitarity formula %%%%%%%%%%%%%%
\begin{eqnarray}
\label{eq:unitary}
\frac{\Delta\Sigma^{{\rm PQ}}_\nu(x,\{z_i\})}{\Sigma}
+1+\frac{\nu^2}{x^2}=\hspace{2in}\nonumber\\
\sum_{j=1}^{N_f}\frac{2}{x^2-z_j^2}\left[
\frac{x\Sigma^{{\rm PQ}}_\nu(x,\{z_i\})}{\Sigma}
-\frac{z_j\Sigma^{{\rm full}(N_f,j)}_\nu(\{z_i\})}{\Sigma}
\right].
\end{eqnarray}
It is not difficult to show that 
the left hand side of Eq.(\ref{eq:unitary}) 
actually vanishes in the quenched limit, {\em cf.} 
Eq.(\ref{eq:DeltaSigmaquenchlimit}).

It is also interesting to consider the limit of the full $N_f$-flavor theory
of  Eq.(\ref{eq:unitary}),
in which all the valence and sea quarks are
degenerate: $x=z_1=z_2=\cdots =z$.
In that degenerate limit we have
\begin{eqnarray}
\label{eq:Sigma-full-deg}
%%%%%%%%%%% Eq full degenerate Sigma %%%%%%%%%%%
\frac{\Sigma^{N_f}_\nu(z)}{\Sigma}
&\equiv& 
 \frac{1}{N_f} \sum_i 
\left.\frac{\Sigma^{{\rm full} (N_f,i)}_\nu(z_1,z_2,\cdots)}{\Sigma}
\right|_{z_1=z_2=\cdots =z},\\
%%%%%%%%%% Eq full degenerate Sigma prime %%%%%
\left(\frac{\Sigma^{N_f}_\nu(z)}{\Sigma}\right)^\prime
%&\equiv& \frac{1}{N_f} \sum_i 
%\frac{\partial_{z_i}\Sigma^{{\rm full} (N_f, i)}_\nu(z,z,\cdots)}
%{\Sigma}
% +\frac{1}{N_f} \sum_{j\neq i} 
%\frac{\partial_{z_j}\Sigma^{{\rm full} (N_f, i)}_\nu(z,z,\cdots)}
%{\Sigma}\nonumber\\
&=& \left.\frac{\partial_{z_1}\Sigma^{{\rm full} (N_f, 1)}_\nu(z_1,z_2,\cdots)}
{\Sigma}\right|_{z_1=z_2=\cdots=z} \nonumber\\
&& +\left.(N_f-1)\frac{\partial_{z_2}\Sigma^{{\rm full} (N_f, 1)}_\nu(z_1,z_2,\cdots)}
{\Sigma}\right|_{z_1=z_2=\cdots=z} ~, \nonumber\\
\end{eqnarray}
where we have used the fact that the $\partial_{z_j}\Sigma^{{\rm full} (i)}_\nu$'s
are independent of $i$ or $j$ in the generate case.
We thus have
%%%%%%%%%%% Eq Sigma prime and derivative of Sigma %%%%%%
\begin{eqnarray}
\label{eq:sigmadif1}
\left.\frac{\partial_{z_2}
\Sigma^{{\rm full} (N_f, 1)}_\nu(z_1,z_2,\cdots)}{\Sigma}
\right|_{z_1=z_2=\cdots=z}
=\hspace{2in}\nonumber\\
\frac{1}{N_f-1}
\left(\left(\frac{\Sigma^{N_f}_\nu(z)}{\Sigma}\right)^\prime
-\left.\frac{\partial_{z_1}\Sigma^{{\rm full} (N_f, 1)}_\nu(z_1,z_2,\cdots)}
{\Sigma}\right|_{z_1=z_2=\cdots=z}\right).
\nonumber\\
\end{eqnarray}
%Next, consider the relation between them with the partially quenched 
%condensate.
%We have 
%%%%%%%%%%% Eq PQ Sigma and full Sigma %%%%%%%%%%%%%%%%%%%
%\begin{equation}
%\frac{\Sigma^{{\rm PQ}}_\nu(x\to z_i+\epsilon, 
%\{\cdots z_i+\epsilon,\cdots \})}{\Sigma}
%=\frac{\Sigma^{{\rm full} (N_f, i)}_\nu (\cdots z_i+\epsilon,\cdots )}{\Sigma},
%\end{equation}
%and it follows from this equation that
%%%%%%%%%%% Eq PQ Sigma and full Sigma derivative %%%%%%%%%%%
%\begin{eqnarray}
%\label{eq:sigmadif2}
%\hspace{-0.2in}
%\frac{\partial_x\Sigma^{{\rm PQ}}_\nu(z_i, 
%\{\mbox{\tiny $\cdots$} z_i, \mbox{\tiny $\cdots$} \})}{\Sigma}
%+\frac{\partial_{z_i}\Sigma^{{\rm PQ}}_\nu(z_i, 
%\{\mbox{\tiny $\cdots$} z_i,\mbox{\tiny $\cdots$} \})}{\Sigma}
%=\frac{\partial_{z_i}\Sigma^{{\rm full} (N_f, i)}_\nu 
%(\mbox{\tiny $\cdots$} z_i,\mbox{\tiny $\cdots$} )}{\Sigma}.
%\end{eqnarray}
%Also, we know
%%%%%%%%%%%% Eq PQ Sigma and full Sigma 2 %%%%%%%%%%%%%%%%%%%
%\begin{eqnarray}
%\frac{\Sigma^{{\rm PQ}}_\nu(z,
%\{z, z, \mbox{\tiny$\cdots$} z_i=z+\epsilon,\mbox{\tiny$\cdots$} \})}{\Sigma}
%=\frac{\Sigma^{{\rm full} (N_f, j)}_\nu (z, z,\mbox{\tiny$\cdots$}
%z_i=z+\epsilon,\mbox{\tiny$\cdots$} )}
%{\Sigma}\;\;(i\neq j),
%\nonumber\\
%\end{eqnarray}
%which leads to 
%%%%%%%%%%%%% Eq PQ Sigma full Sigma derivative 2 %%%%%%%%%%%
%\begin{eqnarray}
%\label{eq:sigmadif3}
%\frac{\partial_{z_j}\Sigma^{{\rm PQ}}_\nu(z,
%\{z, z, \cdots \})}{\Sigma}
%=\frac{\partial_{z_2}\Sigma^{{\rm full} (N_f, 1)}_\nu (z, z,\cdots)}
%{\Sigma}.
%\end{eqnarray}
%From Eqs.(\ref{eq:sigmadif1}), (\ref{eq:sigmadif2}) and 
%(\ref{eq:sigmadif3}), an equation
Similarly, one obtains
%%%%%%%%%%%%% Eq PQ Sigma derivative, full, prime %%%%%%%%%%%%
\begin{eqnarray}
\label{eq:sigmadif4}
\left.\frac{\partial_{z_1}\Sigma^{{\rm full} (N_f, 1)}_\nu 
( z_1, z_2,\mbox{\tiny $\cdots$})}{\Sigma}\right|_{z_1=z_2=\cdots=z}
=\hspace{2in}\nonumber\\
\left.\frac{N_f-1}{N_f}\frac{\partial_x\Sigma^{{\rm PQ}}_\nu(x, 
\{z, z,\mbox{\tiny $\cdots$} \})}{\Sigma}\right|_{x=z}
+\frac{1}{N_f}\left(\frac{\Sigma^{N_f}_\nu(z)}
{\Sigma}\right)^\prime. \nonumber\\
\end{eqnarray}

The unitarity equation
for the degenerate case ($x=z_1=z_2=\cdots =z$), 
Eq.(\ref{eq:unitary}) then finally becomes
%%%%%%%%%%%%% Eq Unitarity formula full theory %%%%%%%%%%%%%%%%
\begin{eqnarray}
\label{eq:unitary-full}
-\left(\frac{\Sigma^{N_f}_\nu(z)}{\Sigma}\right)^2
-\frac{1}{N_f}\left(\frac{\Sigma^{N_f}_\nu(z)}{\Sigma}\right)^\prime
-N_f\frac{\Sigma^{N_f}_\nu(z)}{z\Sigma}
+1+\frac{\nu^2}{z^2}
\nonumber\\
&&\hspace{-3in}=\frac{(N_f^2-1)}{N_f}
\left.\frac{\partial_x\Sigma^{{\rm PQ}}_\nu(x,\{z,z,\cdots\})}{\Sigma}\right|_{x=z}\nonumber\\
&&\hspace{-3in}=(N_f+1)\left(\left.\frac{\partial_{z_1}\Sigma^{{\rm full} (N_f, 1)}_\nu 
(\{ z_1, z,z,\mbox{\tiny $\cdots$}\})}{\Sigma}\right|_{z_1=z}
-\frac{\Sigma^{\prime}_\nu(z)}
{N_f\Sigma}
\right) ~,
\end{eqnarray}
a relation which is useful when simplifying the expressions
for the meson correlators in 
Sec.\ref{sec:meson}.
We finally note that  
Eq.(\ref{eq:unitary-full}) is consistent
with an analogous formula obtained 
from Schwinger-Dyson equations \cite{Damgaard:2001js} in the full
theory.

%%%%%%%%%%%%%%%%%%%%%%%%%%%%%%%%%%%%%%%%%%%%%%%%%%%%%%%
%%%%%%%%%%%%%%%%%%%%%%%%%%%%%%%%%%%%%%%%%%%%%%%%%%%%%%%
\section{Meson correlators}
\label{sec:meson}
\setcounter{equation}{0}
%%%%%%%%%%%%%%%%%%%%%%%%%%%%%%%%%%%%%%%%%%%%%%%%%%%%%%%
%%%%%%%%%%%%%%%%%%%%%%%%%%%%%%%%%%%%%%%%%%%%%%%%%%%%%%%

In this section we present the detailed analytical predictions
of partially quenched scalar and pseudoscalar correlation functions 
in the $\epsilon$-regime of QCD. Our calculation is done
to the lowest non-trivial order in the $\epsilon$-expansion, 
and by taking limits we can
check that we recover both the fully quenched and $N_f=2$
results at sectors of fixed topological index $\nu$ as
reported in ref. \cite{Damgaard:2001js}.
 
%%%%%%%%%%%%%%%%%%%%%%%%%%%%%%%%%%%%%%%%%%%%%%%%%%%%%%%
\subsection{Partially Quenched Correlators with Two Valence Quarks}
%%%%%%%%%%%%%%%%%%%%%%%%%%%%%%%%%%%%%%%%%%%%%%%%%%%%%%%

Of most immediate interest
are the correlation functions
with two light valence quarks, representing the $u$ and $d$ quarks
in QCD. The physical case of low-energy QCD with two 
very light quarks 
(the $u$ and the $d$ quarks) and one heavier quark (the $s$
quark) is an example where one can be in a mixture of the
$\epsilon$-regime (with respect to the two light quarks)  
and the $p$-regime (with respect to the $s$ quark). Even if
the two light quarks correspond, for the given lattice volume,
to the $\epsilon$-regime, it may be convenient to recycle
the lattice configurations by considering light valence
quarks that are {\em also} in the $\epsilon$-regime, but
just at different mass values. This requires a comparison
with the results of PQChPT in the $\epsilon$-regime that we shall
present here.

Furthermore,
with the resulting formulae as building blocks one can calculate 
various other different types of meson correlators. They can
correspond not just to different valence
and sea quark masses but also to an
arbitrary number of valence and sea quark flavors.
For example, the $SU(3)$-singlet scalar correlator is obtained by
%%%%%%%%%% Eq example SU(3) singlet %%%%%%%%%%%%%%%%%%%%%
\begin{equation}
\langle S^0(x)S^0(0) \rangle =
\sum^3_i\sum^3_j  
\langle \bar{q}_{v_i}q_{v_i}(x)\bar{q}_{v_j}q_{v_j}(0)\rangle_d
+\sum^3_i \langle \bar{q}_{v_i}q_{v_i}(x)\bar{q}_{v_i}q_{v_i}(0)\rangle_c,
\end{equation}
where the expressions in the r.h.s. are defined below.
Note that the contributions inside the sums can differ 
due to the non-degeneracy of the valence quark masses.

The connected scalar and pseudo-scalar correlators 
are defined by
\begin{eqnarray}
%%%%%%%%%%%% Eq connected SS %%%%%%%%%%%%%%%%%%%%
\label{eq:conS}
\langle \bar{q}_{v_1}q_{v_2}(x)\bar{q}_{v_2}q_{v_1}(0)\rangle_c
\hspace{3in}\nonumber\\
\equiv
\frac{\Sigma^2}{8}
\left\langle\left(U(x)_{v_1 v_2}+U^\dagger(x)_{v_2 v_1}+
U(x)_{v_2 v_1}+U^\dagger(x)_{v_1 v_2}\right)\right.
\nonumber\\
\left.\times\left(U(0)_{v_1 v_2}+U^\dagger(0)_{v_2 v_1}+
U(0)_{v_2 v_1}+U^\dagger(0)_{v_1 v_2}\right)\right\rangle,
\nonumber\\\\
%%%%%%%%%%% Eq connected PP %%%%%%%%%%%%%%%%%%%%%
\label{eq:conP}
\langle \bar{q}_{v_1}\gamma_5q_{v_2}(x)
\bar{q}_{v_2}\gamma_5q_{v_1}(0)\rangle_c
\hspace{3in}\nonumber\\
\equiv
\frac{\Sigma^2}{8}
\left\langle\left(U(x)_{v_1 v_2}-U^\dagger(x)_{v_2 v_1}+
U(x)_{v_2 v_1}-U^\dagger(x)_{v_1 v_2}\right)\right.\nonumber\\
\left.\times\left(U(0)_{v_1 v_2}-U^\dagger(0)_{v_2 v_1}+
U(0)_{v_2 v_1}-U^\dagger(0)_{v_1 v_2}\right)\right\rangle,\nonumber\\
\end{eqnarray}
where $v_i$ denotes the valence flavor index with the quark mass 
$m_{v_i}=x_i/\Sigma V$.
Note that Eq.(\ref{eq:conS}) and Eq.(\ref{eq:conP}) have the same structure
as the iso-triplet correlators in the 2-flavor theory
of which masses are $m_{v_1}$ and $m_{v_2}$.

The disconnected correlators are similarly defined
%%%%%%%%%%% Eq disconnected SS %%%%%%%%%%%%%%%
\begin{eqnarray}
\langle \bar{q}_{v_1}q_{v_1}(x)\bar{q}_{v_2}q_{v_2}(0)\rangle_d
\hspace{3in}\nonumber\\
\equiv
\frac{\Sigma^2}{4}
\left\langle\left(U(x)_{v_1 v_1}+U^\dagger(x)_{v_1 v_1}\right)
\left(U(0)_{v_2 v_2}+U^\dagger(0)_{v_2 v_2}\right)\right\rangle,
\nonumber\\\\\
\hspace{-0.5in}
%%%%%%%%%%% Eq disconnected PP %%%%%%%%%%%%%%%
\langle \bar{q}_{v_1}\gamma_5q_{v_1}(x)
\bar{q}_{v_2}\gamma_5q_{v_2}(0)\rangle_d
\hspace{3in}\nonumber\\
\equiv
\frac{\Sigma^2}{4}
\left\langle\left(U(x)_{v_1 v_1}-U^\dagger(x)_{v_1 v_1}\right)
\left(U(0)_{v_2 v_2}-U^\dagger(0)_{v_2 v_2}\right)\right\rangle.\nonumber\\
\end{eqnarray}

Let us begin with the disconnected scalar correlators.
To $O(\epsilon^2)$, we find
%%%%%%%%%%%% Eq dis SS with U %%%%%%%%%%%%%%%%
\begin{eqnarray}
\langle \bar{q}_{v_1}q_{v_1}(x)\bar{q}_{v_2}q_{v_2}(0)\rangle_d
\nonumber\\
&&
\hspace{-1.2in}=\frac{\Sigma^2}{4}
\left\langle (U_{v_1 v_1}+U^\dagger_{v_1 v_1})(U_{v_2 v_2}+U^\dagger_{v_2 v_2})
\right\rangle^{\rm 1-loop}\times
%\left(\frac{\Sigma^{v_1}_{\rm eff}}{\Sigma}\right)
%\left(\frac{\Sigma^{v_2}_{\rm eff}}{\Sigma}\right)
\left(\frac{x_1^{\rm eff}}{x_1}\right)
\left(\frac{x_2^{\rm eff}}{x_2}\right)
%\left(1-\frac{1}{F^2}\sum^{N+N_f}_i\bar{P}_{(2i)(i2)}(0)\right)^2
\nonumber\\
&&\hspace{-1.0in}-\frac{\Sigma^2}{2F^2}
\langle U_{v_1v_2}U_{v_2v_1}+U^\dagger_{v_2v_1}U^\dagger_{v_1v_2}\rangle
\bar{P}_{(12)(21)}(x)\nonumber\\
&&\hspace{-1.0in}-\frac{\Sigma^2}{2F^2}\langle (U_{v_1v_1}-U^\dagger_{v_1v_1})
(U_{v_2v_2}-U^\dagger_{v_2v_2})\rangle\bar{P}_{(11)(22)}(x)\nonumber\\
&&\hspace{-1.2in}=\frac{\Sigma^2}{4}\left(\frac{x_1^{\rm eff}}{x_1}\right)^2
\left\langle (U_{v_1 v_1}+U^\dagger_{v_1 v_1})(U_{v_2 v_2}+U^\dagger_{v_2 v_2})
\right\rangle^{\rm 1-loop}\nonumber\\
&&\hspace{-1.0in}-\frac{\Sigma^2}{2F^2}
\langle U_{v_1v_2}U_{v_2v_1}+U^\dagger_{v_2v_1}U^\dagger_{v_1v_2}\rangle
\bar{\Delta}(0|x)\nonumber\\
&&\hspace{-1.0in}-\frac{\Sigma^2}{2F^2}\langle (U_{v_1v_1}-U^\dagger_{v_1v_1})
(U_{v_2v_2}-U^\dagger_{v_2v_2})\rangle\bar{G}(0,0|x),
\end{eqnarray}
We have consistently set $m_{v_1}=m_{v_2}=0$ in the NLO
contributions, and
%%%%%%%%%% Eq x_eff %%%%%%%%%%%%%%%%%%%%%%%%%%%%%%% 
\begin{equation}
\frac{x_1^{\rm eff}}{x_1}= \frac{x_2^{\rm eff}}{x_2}
=1-\frac{1}{F^2}\left(
\sum_i^{N_f}\bar{\Delta}(M^2_{i v}|0)
-\bar{G}(0,0|0)\right),
\end{equation}
which of course needs regularization.
 Note $\langle \cdots \rangle^{\rm 1-loop}$ indicates the
shift $x_i\to x_i^{\rm eff}$
and $z_i\to z^{\rm eff}_i$
in the arguments of the Bessel functions.

In the same way we obtain the disconnected pseudo-scalar correlation function
%%%%%%%%%%% Eq dis PP with U %%%%%%%%%%%%%%%%%%%%%%%
\begin{eqnarray}
\langle \bar{q}_{v_1}\gamma_5 q_{v_1}(x)\bar{q}_{v_2}\gamma_5 q_{v_2}(0)\rangle_d
\hspace{-1in}\nonumber\\
&=&\frac{\Sigma^2}{4}\left(\frac{x_1^{\rm eff}}{x_1}\right)^2
\left\langle (U_{v_1 v_1}-U^\dagger_{v_1 v_1})(U_{v_2 v_2}-U^\dagger_{v_2 v_2})
\right\rangle^{\rm 1-loop}\nonumber\\
&&-\frac{\Sigma^2}{2F^2}
\langle U_{v_1v_2}U_{v_2v_1}+U^\dagger_{v_2v_1}U^\dagger_{v_1v_2}\rangle
\bar{\Delta}(0|x)\nonumber\\
&&-\frac{\Sigma^2}{2F^2}\langle (U_{v_1v_1}+U^\dagger_{v_1v_1})
(U_{v_2v_2}+U^\dagger_{v_2v_2})\rangle\bar{G}(0,0|x).
\nonumber\\
\end{eqnarray}

For the connected correlators, we get
%%%%%%%%%%%% Eq con SS with U %%%%%%%%%%%%%%%%%%%%%
\begin{eqnarray}
\langle \bar{q}_{v_1}q_{v_2}(x)\bar{q}_{v_2}q_{v_1}(0)\rangle_c
\hspace{-1.3in}\nonumber\\
&
=&\frac{\Sigma^2}{4}\left(\frac{x_1^{\rm eff}}{x_1}\right)^2
\left\langle (U_{v_1 v_2}+U^\dagger_{v_2 v_1})^2+
(U_{v_1 v_2}+U^\dagger_{v_2 v_1})(U_{v_2 v_1}+U^\dagger_{v_1 v_2})
\right\rangle^{\rm 1-loop}\nonumber\\
&&
+\frac{\Sigma^2}{2F^2}\sum_i^\prime
\left[\langle U_{v_1 i}U^\dagger_{i v_1}\rangle\bar{P}_{(2i)(i2)}(x)
+\langle U_{v_2 i}U^\dagger_{i v_2}\rangle\bar{P}_{(1i)(i1)}(x)
\right]\nonumber\\
&&
-\frac{\Sigma^2}{4F^2}\left[
\langle U_{v_1v_2}^2+(U^\dagger_{v_2v_1})^2\rangle
\bar{P}_{(22)(22)}(x)
+\langle U_{v_2v_1}^2+(U^\dagger_{v_1v_2})^2\rangle
\bar{P}_{(11)(11)}(x)\right]
\nonumber\\
&&
-\frac{\Sigma^2}{4F^2}
\langle U_{v_1v_1}U_{v_2 v_2}+U^\dagger_{v_1v_1}U^\dagger_{v_2v_2}\rangle
\left[\bar{P}_{(12)(21)}(x)+\bar{P}_{(21)(12)}(x)\right]
\nonumber\\
&&
-\frac{\Sigma^2}{4F^2}
\langle (U_{v_1v_2}-U^\dagger_{v_2v_1})(U_{v_2v_1}-U^\dagger_{v_1v_2})\rangle
\bar{P}_{(22)(11)}(x)
\nonumber\\
&&
-\frac{\Sigma^2}{4F^2}
\langle (U_{v_2v_1}-U^\dagger_{v_1v_2})(U_{v_1v_2}-U^\dagger_{v_2v_1})\rangle
\bar{P}_{(11)(22)}(x),
\end{eqnarray}
%%%%%%%%%%%%%%% Eq con PP with U %%%%%%%%%%%%%%%%%%
\begin{eqnarray}
\langle \bar{q}_{v_1}\gamma_5 q_{v_2}(x)\bar{q}_{v_2}\gamma_5 q_{v_1}(0)\rangle_c
\hspace{-1.5in}\nonumber\\
&=&\frac{\Sigma^2}{4}\left(\frac{x_1^{\rm eff}}{x_1}\right)^2
\left\langle (U_{v_1 v_2}-U^\dagger_{v_2 v_1})^2+
(U_{v_1 v_2}-U^\dagger_{v_2 v_1})(U_{v_2 v_1}-U^\dagger_{v_1 v_2})
\right\rangle^{\rm 1-loop}\nonumber\\
&&
-\frac{\Sigma^2}{2F^2}\sum_i^\prime
\left[\langle U_{v_1 i}U^\dagger_{i v_1}\rangle\bar{P}_{(2i)(i2)}(x)
+\langle U_{v_2 i}U^\dagger_{i v_2}\rangle\bar{P}_{(1i)(i1)}(x)
\right]\nonumber\\
&&
-\frac{\Sigma^2}{4F^2}\left[
\langle U_{v_1v_2}^2+(U^\dagger_{v_2v_1})^2\rangle
\bar{P}_{(22)(22)}(x)
+\langle U_{v_2v_1}^2+(U^\dagger_{v_1v_2})^2\rangle
\bar{P}_{(11)(11)}(x)\right]
\nonumber\\
&&
-\frac{\Sigma^2}{4F^2}
\langle U_{v_1v_1}U_{v_2 v_2}+U^\dagger_{v_1v_1}U^\dagger_{v_2v_2}\rangle
\left[\bar{P}_{(12)(21)}(x)+\bar{P}_{(21)(12)}(x)\right]
\nonumber\\
&&
-\frac{\Sigma^2}{4F^2}
\langle (U_{v_1v_2}+U^\dagger_{v_2v_1})(U_{v_2v_1}+U^\dagger_{v_1v_2})\rangle
\bar{P}_{(22)(11)}(x)
\nonumber\\
&&
-\frac{\Sigma^2}{4F^2}
\langle (U_{v_2v_1}+U^\dagger_{v_1v_2})(U_{v_1v_2}+U^\dagger_{v_2v_1})\rangle
\bar{P}_{(11)(22)}(x),
\end{eqnarray}
where the summation over the flavors, 
denoted by $\sum_i^\prime$, has to be taken carefully.
For example, in the sum
%%%%%%%%%%%%% Eq sum UU %%%%%%%%%%%%%%%%%%%%%%%%%%%
\begin{eqnarray}
\sum_i^\prime
\langle U_{v_1 i}U^\dagger_{i v_1}\rangle\bar{P}_{(2i)(i2)}(x)
\nonumber\\
&&\hspace{-1.2in}\equiv \lim_{N_{v_1}\to 0}\sum_i^{N_{v_1}}  
\langle U_{v_1 i}U^\dagger_{i v_1}\rangle\bar{P}_{(2i)(i2)}(x)
+\lim_{N_{v_2}\to 0}\sum_i^{N_{v_2}}  
\langle U_{v_1 i}U^\dagger_{i v_1}\rangle\bar{P}_{(2i)(i2)}(x)
\nonumber\\
&&\hspace{-1.0in}+\sum_i^{N_f}  
\langle U_{v_1 i}U^\dagger_{i v_1}\rangle\bar{P}_{(2i)(i2)}(x)\nonumber\\
&&\hspace{-1.2in}=\langle U_{v_1 v_1}U^\dagger_{v_1 v_1}\rangle\bar{P}_{(2 1)(1 2)}(x)
- \langle U_{v_1 v^\prime_1}U^\dagger_{v^\prime_1 v_1}\rangle
\bar{P}_{(2 1)(1 2)}(x)\nonumber\\
&&\hspace{-1.0in}+\langle U_{v_1 v_2}U^\dagger_{v_2 v_1}\rangle\bar{P}_{(2 2)(2 2)}(x)
- \langle U_{v_1 v_2}U^\dagger_{v_2 v_1}\rangle
\bar{P}_{(2 2^\prime)(2^\prime 2)}(x)\nonumber\\
&&+\sum_i^{N_f}  
\langle U_{v_1 i}U^\dagger_{i v_1}\rangle\bar{P}_{(2i)(i2)}(x),
\end{eqnarray}
the index with a prime $i^\prime$ is treated as
a different flavor from $i$-th quark 
but with the same value of the quark mass.
With this, we find the correlators
%%%%%%%%%%%%%% Eq con SS with U 2 %%%%%%%%%%%%%%%%%%%
\begin{eqnarray}
\langle \bar{q}_{v_1}q_{v_2}(x)\bar{q}_{v_2}q_{v_1}(0)\rangle_c
\hspace{-1.5in}\nonumber\\
&&\hspace{-0.2in}
=\frac{\Sigma^2}{4}
\left(\frac{x_1^{\rm eff}}{x_1}\right)^2
\left\langle (U_{v_1 v_2}+U^\dagger_{v_2 v_1})^2+
(U_{v_1 v_2}+U^\dagger_{v_2 v_1})(U_{v_2 v_1}+U^\dagger_{v_1 v_2})
\right\rangle^{\rm 1-loop}\nonumber\\
&&
+\frac{\Sigma^2}{2F^2}
\langle U_{v_1 v_1}U^\dagger_{v_1 v_1}-U_{v_1 v^\prime_1}U^\dagger_{v^\prime_1 v_1}
+U_{v_2 v_2}U^\dagger_{v_2 v_2}-U_{v_2 v^\prime_2}U^\dagger_{v^\prime_2 v_2}
\nonumber\\
&&
\hspace{2in}-U_{v_1 v_1}U_{v_2 v_2}-U^\dagger_{v_1 v_1}U^\dagger_{v_2 v_2}
\rangle\bar{\Delta}(0|x)
\nonumber\\
&&-\frac{\Sigma^2}{2F^2}
\langle U_{v_1v_2}U^\dagger_{v_2v_1}+U_{v_2v_1}U^\dagger_{v_1v_2}
%\nonumber\\
%&&
-(U_{v_1v_2}-U^\dagger_{v_2v_1})(U_{v_2v_1}-U^\dagger_{v_1v_2})
\rangle
\bar{G}(0,0|x)
\nonumber\\
&&
+\frac{\Sigma^2}{2F^2}\sum_i^{N_f}
\langle U_{v_1 i}U^\dagger_{i v_1}+U_{v_2 i}U^\dagger_{i v_2}\rangle
\bar{\Delta}(M^2_{iv}|x),
\end{eqnarray}
%%%%%%%%%%%% Eq con PP with U 2   %%%%%%%%%%%%%%%
\begin{eqnarray}
\langle \bar{q}_{v_1}\gamma_5 q_{v_2}(x)\bar{q}_{v_2}\gamma_5 q_{v_1}(0)\rangle_c
\hspace{-1.7in}\nonumber\\
&&\hspace{-0.2in}
=\frac{\Sigma^2}{4}\left(\frac{x_1^{\rm eff}}{x_1}\right)^2
\left\langle (U_{v_1 v_2}-U^\dagger_{v_2 v_1})^2+
(U_{v_1 v_2}-U^\dagger_{v_2 v_1})(U_{v_2 v_1}-U^\dagger_{v_1 v_2})
\right\rangle^{\rm 1-loop}\nonumber\\
&&
-\frac{\Sigma^2}{2F^2}
\langle U_{v_1 v_1}U^\dagger_{v_1 v_1}-U_{v_1 v^\prime_1}U^\dagger_{v^\prime_1 v_1}
+U_{v_2 v_2}U^\dagger_{v_2 v_2}-U_{v_2 v^\prime_2}U^\dagger_{v^\prime_2 v_2}
\nonumber\\
&&
\hspace{2in}+U_{v_1 v_1}U_{v_2 v_2}+U^\dagger_{v_1 v_1}U^\dagger_{v_2 v_2}
\rangle\bar{\Delta}(0|x)\nonumber\\
&&
+\frac{\Sigma^2}{2F^2}
\langle U_{v_1v_2}U^\dagger_{v_2v_1}+U_{v_2v_1}U^\dagger_{v_1v_2}
%\nonumber\\
%&&
%\hspace{2in}
+(U_{v_1v_2}+U^\dagger_{v_2v_1})(U_{v_2v_1}+U^\dagger_{v_1v_2})
\rangle
\bar{G}(0,0|x)
\nonumber\\
&&
-\frac{\Sigma^2}{2F^2}\sum_i^{N_f}
\langle U_{v_1 i}U^\dagger_{i v_1}+U_{v_2 i}U^\dagger_{i v_2}\rangle
\bar{\Delta}(M^2_{iv}|x),
\end{eqnarray}
where we have explicitly kept the sea quark mass, in order to possibly apply these
equations outside of the the $\epsilon$-regime.\\

%\newpage
With aid of the formulae collected in Appendix \ref{app:Uint}
and use of the defining equation (\ref{eq:DPQ}) for $D_\nu^{{\rm PQ}}$, 
one obtains
%%%%%%%%%%%% Eq dis SS with Sigma %%%%%%%%%%%%%%%%%%%%%
\begin{eqnarray}
\langle \bar{q}_{v_1}q_{v_1}(x)\bar{q}_{v_2}q_{v_2}(0)\rangle_d
&=&
\Sigma^2\left(\frac{x_1^{\rm eff}}{x_1}\right)^2D_\nu^{{\rm PQ}}(x^{\rm eff}_1, x^{\rm eff}_2, \{z^{\rm eff}_i\})
\nonumber\\
&&\hspace{-1in}-\frac{\Sigma^2}{F^2}
\frac{2}{x^2_1-x^2_2}\left(\frac{x_2\Sigma_{\nu}^{{\rm PQ}}(x_1, \{z_i\})}{\Sigma}
-\frac{x_1\Sigma_{\nu}^{{\rm PQ}}(x_2, \{z_i\})}{\Sigma}
\right)\bar{\Delta}(0|x)\nonumber\\
&&\hspace{-1in}+\frac{\Sigma^2}{F^2}\frac{2\nu^2}{x_1x_2}\bar{G}(0,0|x),
\end{eqnarray}
%%%%%%%%%%%% Eq dis PP with Sigma %%%%%%%%%%%%%%%%%%%%%
\begin{eqnarray}
\langle \bar{q}_{v_1}\gamma_5 q_{v_1}(x)\bar{q}_{v_2}\gamma_5q_{v_2}(0)\rangle_d
&=&
\Sigma^2\left(\frac{x_1^{\rm eff}}{x_1}\right)^2\frac{\nu^2}{x_1^{\rm eff}x_2^{\rm eff}}
\nonumber\\
&&\hspace{-1in}-\frac{\Sigma^2}{F^2}
\frac{2}{x^2_1-x^2_2}\left(\frac{x_2\Sigma_{\nu}^{{\rm PQ}}(x_1, \{z_i\})}{\Sigma}
-\frac{x_1\Sigma_{\nu}^{{\rm PQ}}(x_2, \{z_i\})}{\Sigma}
\right)\bar{\Delta}(0|x)\nonumber\\
&&\hspace{-1in}
+\frac{2\Sigma^2}{F^2}D_\nu^{{\rm PQ}}(x_1, x_2, \{z_i\})\bar{G}(0,0|x),
\end{eqnarray}
%%%%%%%%%%%% Eq con SS with Sigma %%%%%%%%%%%%%%%%%%%%%
\begin{eqnarray}
\langle \bar{q}_{v_1}q_{v_2}(x)\bar{q}_{v_2}q_{v_1}(0)\rangle_c=
\nonumber\\
&&\hspace{-1.5in}
\Sigma^2\left(\frac{x_1^{\rm eff}}{x_1}\right)^2
\left[
\frac{1}{x_1^{\rm eff}-x_2^{\rm eff}}
\left(
\frac{\Sigma_{\nu}^{{\rm PQ}}(x_1^{\rm eff},\{z_i^{\rm eff}\})}{\Sigma}
-\frac{\Sigma_{\nu}^{{\rm PQ}}(x_2^{\rm eff},\{z_i^{\rm eff}\})}{\Sigma}
\right)
\right]
\nonumber\\
&&\hspace{-1.5in}-\frac{\Sigma^2}{2F^2}
\left(
\frac{\Delta\Sigma_{\nu}^{{\rm PQ}}(x_1,\{z_i\})}{\Sigma}
+\frac{\Delta\Sigma_{\nu}^{{\rm PQ}}(x_2,\{z_i\})}{\Sigma}
+\frac{\nu^2}{x_1^2}+\frac{\nu^2}{x_2^2}\right.
\nonumber\\
&&\hspace{-1.5in}\hspace{1in}\left.+2D_\nu^{{\rm PQ}}(x_1, x_2, \{z_i\})+\frac{2\nu^2}{x_1x_2}
\right)\bar{\Delta}(0|x)\nonumber\\
&&\hspace{-1.5in}-\frac{\Sigma^2}{2F^2}
\left[\frac{4}{x_1+x_2}\left(\frac{\Sigma_{\nu}^{{\rm PQ}}(x_1, \{z_i\})}{\Sigma}
+\frac{\Sigma_{\nu}^{{\rm PQ}}(x_2, \{z_i\})}{\Sigma}
\right)\right]\bar{G}(0,0|x)\nonumber\\
&&\hspace{-1.5in}+\frac{\Sigma^2}{2F^2}\sum_j^{N_f}\left[
\frac{2}{x_1^2-z_j^2}\left(
\frac{x_1\Sigma_{\nu}^{{\rm PQ}}(x_1, \{z_i\})}{\Sigma}
-\frac{z_j\Sigma_{\nu}^{{\rm full}(N_f, j)}(\{z_i\})}{\Sigma}
\right)\right.\nonumber\\
&&\hspace{-1.5in}\left.
+\frac{2}{x_2^2-z_j^2}\left(
\frac{x_2\Sigma_{\nu}^{{\rm PQ}}(x_2, \{z_i\})}{\Sigma}
-\frac{z_j\Sigma_{\nu}^{{\rm full}(N_f, j)}(\{z_i\})}{\Sigma}
\right)
\right]\bar{\Delta}(M^2_{j v}|x),
\end{eqnarray}
%%%%%%%%%%%% Eq con PP with Sigma %%%%%%%%%%%%%%%%%%%%%
\begin{eqnarray}
\langle \bar{q}_{v_1}\gamma_5 q_{v_2}(x)\bar{q}_{v_2}\gamma_5 q_{v_1}(0)\rangle_c
=
\nonumber\\
&&\hspace{-1.5in}
-\Sigma^2\left(\frac{x_1^{\rm eff}}{x_1}\right)^2
\left[
\frac{1}{x_1^{\rm eff}+x_2^{\rm eff}}
\left(
\frac{\Sigma_{\nu}^{{\rm PQ}}(x_1^{\rm eff},\{z_i^{\rm eff}\})}{\Sigma}
+\frac{\Sigma_{\nu}^{{\rm PQ}}(x_2^{\rm eff},\{z_i^{\rm eff}\})}{\Sigma}
\right)
\right]
\nonumber\\
&&\hspace{-1.5in}+\frac{\Sigma^2}{2F^2}
\left(
\frac{\Delta\Sigma_{\nu}^{{\rm PQ}}(x_1,\{z_i\})}{\Sigma}
+\frac{\Delta\Sigma_{\nu}^{{\rm PQ}}(x_2,\{z_i\})}{\Sigma}
+\frac{\nu^2}{x_1^2}+\frac{\nu^2}{x_2^2}\right.
\nonumber\\
&&\hspace{-1.5in}\left.-2D_\nu^{{\rm PQ}}(x_1, x_2, \{z_i\})-\frac{2\nu^2}{x_1x_2}
\right)\bar{\Delta}(0|x)\nonumber\\
&&\hspace{-1.5in}+\frac{\Sigma^2}{2F^2}
\left[\frac{4}{x_1-x_2}\left(\frac{\Sigma_{\nu}^{{\rm PQ}}(x_1, \{z_i\})}{\Sigma}
-\frac{\Sigma_{\nu}^{{\rm PQ}}(x_2, \{z_i\})}{\Sigma}
\right)\right]\bar{G}(0,0|x)\nonumber\\
&&\hspace{-1.5in}-\frac{\Sigma^2}{2F^2}\sum_j^{N_f}\left[
\frac{2}{x_1^2-z_j^2}\left(
\frac{x_1\Sigma_{\nu}^{{\rm PQ}}(x_1, \{z_i\})}{\Sigma}
-\frac{z_j\Sigma_{\nu}^{{\rm full}(N_f, j)}(\{z_i\})}{\Sigma}
\right)\right.\nonumber\\
&&\hspace{-1.5in}\left.
+\frac{2}{x_2^2-z_j^2}\left(
\frac{x_2\Sigma_{\nu}^{{\rm PQ}}(x_2, \{z_i\})}{\Sigma}
-\frac{z_j\Sigma_{\nu}^{{\rm full}(N_f, j)}(\{z_i\})}{\Sigma}
\right)
\right]\bar{\Delta}(M^2_{j v}|x). \nonumber\\
\end{eqnarray}
We have here three types of $\xi$-correlators,
%%%%%%%%%%%% three xi propagators %%%%%%%%%%%%%%
\begin{eqnarray}
\bar{\Delta}(0|x)&=&\frac{1}{V}\sum_{p\neq 0}\frac{e^{ipx}}{p^2},\\
\bar{\Delta}(M_{i v}|x)&=&\frac{1}{V}\sum_{p\neq 0}\frac{e^{ipx}}{p^2+z_i/F^2V},\\
\bar{G}(0,0|x)&=&\frac{1}{V}\sum_{p\neq 0}
\frac{e^{ipx}(m_0^2+\alpha p^2)/N_c}{p^4\mathcal{F}(p^2)}.
\end{eqnarray}
Note that if the sea quarks are much smaller than the cut-off of
ChPT, but still in the $p$-regime,
one can take the $m_0\to \infty$ limit. It leads to
%%%%%%%%%%%% Eq xi correlator p-regime limit %%%%%%
\begin{eqnarray}
\bar{G}(0,0|x) &\to& \frac{1}{V}\sum_{p\neq 0}
\frac{e^{ipx}}{p^4\left(\sum_i^{N_f}\frac{1}{p^2+M^2_{i i}}\right)}
\nonumber\\
&=&\frac{1}{N_f}\left(\bar{\Delta}(0|x)+
\sum^{N_f}_i\frac{M^2_{ii}e^{ipx}}{p^4}\right) + O(M^4_{ii}),
\end{eqnarray}
where the well-known double pole contribution appears due to a mismatch 
of the sea and valence quark masses.
Further simplification is possible when all the sea quarks
are in the $\epsilon$-regime,
%%%%%%%%%%%% Eq xi correlator e-regime limit %%%%%%%%%%
\begin{equation}
\bar{\Delta}(M_{i v}|x) \to \bar{\Delta}(0|x),\;\;\;
\bar{G}(0,0|x)\to \frac{1}{N_f}\bar{\Delta}(0|x).
\end{equation} 

An interesting special case is the degenerate limit $x_1=x_2$, 
where the above formulae become
%%%%%%%%%%%% Eq dis SS degenerate %%%%%%%%%%%%%%%%%%%
\begin{eqnarray}
\label{eq:Sdis-deg}
\langle \bar{q}_{v_1}q_{v_1}(x)\bar{q}_{v_1}q_{v_1}(0)\rangle_d
&=&
-\Sigma^2\left(\frac{x_1^{\rm eff}}{x_1}\right)^2
\frac{\Delta \Sigma_{\nu}^{{\rm PQ}}(x^{\rm eff}_1, \{z^{\rm eff}_i\})}{\Sigma}
\nonumber\\
&&-\frac{\Sigma^2}{F^2}
\left(\frac{\partial_{x}\Sigma_{\nu}^{{\rm PQ}}(x_1, \{z_i\})}{\Sigma}
-\frac{\Sigma_{\nu}^{{\rm PQ}}(x_1, \{z_i\})}{x_1\Sigma}
\right)\bar{\Delta}(0|x)\nonumber\\
&&+\frac{\Sigma^2}{F^2}\frac{2\nu^2}{x_1^2}\bar{G}(0,0|x),
\end{eqnarray}
%%%%%%%%%%%% Eq dis PP degenerate %%%%%%%%%%%%%%%%%%%
\begin{eqnarray}
\label{eq:Pdis-deg}
\langle \bar{q}_{v_1}\gamma_5 q_{v_1}(x)\bar{q}_{v_1}\gamma_5q_{v_1}(0)\rangle_d
&=&
\Sigma^2\left(\frac{x_1^{\rm eff}}{x_1}\right)^2
\frac{\nu^2}{(x_1^{\rm eff})^2}
\nonumber\\
&&-\frac{\Sigma^2}{F^2}
\left(\frac{\partial_x\Sigma_{\nu}^{{\rm PQ}}(x_1, \{z_i\})}{\Sigma}
-\frac{\Sigma_{\nu}^{{\rm PQ}}(x_1, \{z_i\})}{x_1\Sigma}
\right)\bar{\Delta}(0|x)\nonumber\\
&&-\frac{2\Sigma^2}{F^2}
\frac{\Delta \Sigma_{\nu}^{{\rm PQ}}(x_1, \{z_i\})}{\Sigma}\bar{G}(0,0|x),
\end{eqnarray}
%%%%%%%%%%%% Eq con SS degenerate %%%%%%%%%%%%%%%%%%%
\begin{eqnarray}
\label{eq:Scon-deg}
\langle \bar{q}_{v_1}q_{v_1}(x)\bar{q}_{v_1}q_{v_1}(0)\rangle_c=
\nonumber\\
&&\hspace{-1.4in}
\Sigma^2\left(\frac{x_1^{\rm eff}}{x_1}\right)^2
\frac{\partial_x\Sigma_{\nu}^{{\rm PQ}}(x_1^{\rm eff},\{z_i^{\rm eff}\})}{\Sigma}
-\frac{2\Sigma^2}{F^2}\frac{\nu^2}{x_1^2}
\bar{\Delta}(0|x)\nonumber\\
&&\hspace{-1.4in}-\frac{2\Sigma^2}{F^2}
\frac{\Sigma_{\nu}^{{\rm PQ}}(x_1, \{z_i\})}{x_1\Sigma}\bar{G}(0,0|x)\nonumber\\
&&\hspace{-1.4in}+\frac{\Sigma^2}{2F^2}\sum_j^{N_f}\left[
\frac{4}{x_1^2-z_j^2}\left(
\frac{x_1\Sigma_{\nu}^{{\rm PQ}}(x_1, \{z_i\})}{\Sigma}
-\frac{z_j\Sigma_{\nu}^{{\rm full}(N_f, j)}(\{z_i\})}{\Sigma}
\right)
\right]\bar{\Delta}(M^2_{j v}|x),\nonumber\\
\end{eqnarray}
%%%%%%%%%%%% Eq con PP degenerate %%%%%%%%%%%%%%%%%%%
\begin{eqnarray}
\label{eq:Pcon-deg}
\langle \bar{q}_{v_1}\gamma_5 q_{v_1}(x)\bar{q}_{v_1}\gamma_5 q_{v_1}(0)\rangle_c
=\nonumber\\
&&\hspace{-1.9in}
-\Sigma^2\left(\frac{x_1^{\rm eff}}{x_1}\right)^2
\frac{\Sigma_{\nu}^{{\rm PQ}}(x_1^{\rm eff},\{z_i^{\rm eff}\})}{x_1^{\rm eff}\Sigma}
+\frac{2\Sigma^2}{F^2}
\frac{\Delta\Sigma_{\nu}^{{\rm PQ}}(x_1,\{z_i\})}{\Sigma}\bar{\Delta}(0|x)\nonumber\\
&&\hspace{-1.9in}+\frac{2\Sigma^2}{F^2}\frac{\partial_x\Sigma_{\nu}^{{\rm PQ}}(x_1, \{z_i\})}{\Sigma}
\bar{G}(0,0|x)\nonumber\\
&&\hspace{-1.9in}-\frac{\Sigma^2}{2F^2}\sum_j^{N_f}\left[
\frac{4}{x_1^2-z_j^2}\left(
\frac{x_1\Sigma_{\nu}^{{\rm PQ}}(x_1, \{z_i\})}{\Sigma}
-\frac{z_j\Sigma_{\nu}^{{\rm full}(N_f, j)}(\{z_i\})}{\Sigma}
\right)
\right]\bar{\Delta}(M^2_{j v}|x),\nonumber\\
\end{eqnarray}

%%%%%%%%%%%%%%%%%%%%%%%%%%%%%%%%%%%%%%%%%%%%%%%%%%%%%%%
\subsection{Singlet and flavored meson correlators}
%%%%%%%%%%%%%%%%%%%%%%%%%%%%%%%%%%%%%%%%%%%%%%%%%%%%%%%

To compare with lattice QCD data, the case with
arbitrary $N_v$-flavor valence quarks is interesting. In
particular,
$N_v$ can be different from $N_f$, the number of sea quarks.
In this paper we have aimed at the $\epsilon$-regime predictions,
and of course we always take the valence quarks to be
in that regime. With the qualification
mentioned in section 2, one can also be more general, and
consider
the sea quarks to be just marginally in the $\epsilon$-regime,
or even entirely in the $p$-regime.
The finite volume is a $L^3 \times T$ box, where  
$L$ and $T$ are the spacial and temporal extents, respectively.
%We define 
The zero-momentum projections of singlet and 
$(N_v^2-1)$-plet correlators are then obtained 
%%%%%%%%%%%%E Eq singlet SS %%%%%%%%%%%%%%%%%%%%%%%%%%%%%
\begin{eqnarray}
\label{eq:Ssinglet}
\int d^3x \langle S^0(x)S^0(0)\rangle \nonumber\\
&& \hspace{-1.4in}\equiv
\int d^3x \left[N_v \langle \bar{q}_{v_1}q_{v_1}(x)\bar{q}_{v_1}q_{v}(0)\rangle_c
+N_v^2 \langle \bar{q}_{v_1}q_{v_1}(x)\bar{q}_{v_1}q_{v_1}(0)\rangle_d\right]
\nonumber\\
&&\hspace{-1.4in}= L^3\Sigma^2\left(\frac{x_1^{\rm eff}}{x_1}\right)^2\left[N_v
\frac{\partial_x\Sigma_{\nu}^{{\rm PQ}}(x_1^{\rm eff},\{z_i^{\rm eff}\})}{\Sigma}
-N_v^2 \frac{\Delta \Sigma_{\nu}^{{\rm PQ}}(x^{\rm eff}_1, \{z^{\rm eff}_i\})}{\Sigma}
\right]
\nonumber\\
&&
\hspace{-1.3in}-\frac{\Sigma^2}{2F^2}
\left[
\frac{4N_v\nu^2}{x_1^2}
+2N_v^2 \left(\frac{\partial_{x}\Sigma_{\nu}^{{\rm PQ}}(x_1, \{z_i\})}{\Sigma}
-\frac{\Sigma_{\nu}^{{\rm PQ}}(x_1, \{z_i\})}{x_1\Sigma}
\right)
\right]
a(t/T)\nonumber\\
&&\hspace{-1.3in}-\frac{\Sigma^2}{2F^2}
\left[
4N_v\frac{\Sigma_{\nu}^{{\rm PQ}}(x_1, \{z_i\})}{x_1\Sigma}
-4N_v^2 \frac{\nu^2}{x_1^2}
\right]
b(t/T)\nonumber\\
&&\hspace{-1.3in}+\frac{N_v\Sigma^2}{2F^2}\sum_j^{N_f}\left[
\frac{4}{x_1^2-z_j^2}\left(
\frac{x_1\Sigma_{\nu}^{{\rm PQ}}(x_1, \{z_i\})}{\Sigma}
-\frac{z_j\Sigma_{\nu}^{{\rm full}(N_f, j)}(\{z_i\})}{\Sigma}
\right)
\right]c_j(t/T),\nonumber\\
\\
%\end{eqnarray}
%%%%%%%%%%%%%%%%%%% Eq  singlet PP %%%%%%%%%%%%%%
%\begin{eqnarray}
\label{eq:Psinglet}
\int d^3x \langle P^0(x)P^0(0)\rangle &&\nonumber\\
&&\hspace{-1.4in}\equiv 
-\int d^3x \left[N_v\langle 
\bar{q}_{v_1}\gamma_5 q_{v_1}(x)\bar{q}_{v_1}\gamma_5 q_{v_1}(0)\rangle_c
+N_v^2 \langle \bar{q}_{v_1}\gamma_5 q_{v_1}(x)
\bar{q}_{v_1}\gamma_5 q_{v_1}(0)\rangle_d\right]
\nonumber\\
&&\hspace{-1.4in}
=L^3\Sigma^2\left(\frac{x_1^{\rm eff}}{x_1}\right)^2
\left[
N_v\frac{\Sigma_{\nu}^{{\rm PQ}}(x_1^{\rm eff},\{z_i^{\rm eff}\})}{x_1^{\rm eff}\Sigma}
-N_v^2 \frac{\nu^2}{(x_1^{\rm eff})^2}
\right]
\nonumber\\
&&\hspace{-1.3in}
-\frac{\Sigma^2}{2F^2}
\left[
\frac{4N_v\Delta\Sigma_{\nu}^{{\rm PQ}}(x_1,\{z_i\})}{\Sigma}\right.
\nonumber\\
&&\hspace{-0.4in}
\left.-2N_v^2 \left(\frac{\partial_x\Sigma_{\nu}^{{\rm PQ}}(x_1, \{z_i\})}{\Sigma}
-\frac{\Sigma_{\nu}^{{\rm PQ}}(x_1, \{z_i\})}{x_1\Sigma}
\right)
\right]a(t/T)
\nonumber\\
&&\hspace{-1.3in}
-\frac{\Sigma^2}{2F^2}
\left[
4N_v\frac{\partial_x\Sigma_{\nu}^{{\rm PQ}}(x_1, \{z_i\})}{\Sigma}
-4N_v^2 \frac{\Delta \Sigma_{\nu}^{{\rm PQ}}(x_1, \{z_i\})}{\Sigma}
\right]b(t/T)\nonumber\\
&&\hspace{-1.3in}+\frac{N_v\Sigma^2}{2F^2}\sum_j^{N_f}\left[
\frac{4}{x_1^2-z_j^2}\left(
\frac{x_1\Sigma_{\nu}^{{\rm PQ}}(x_1, \{z_i\})}{\Sigma}
-\frac{z_j\Sigma_{\nu}^{{\rm full}(N_f, j)}(\{z_i\})}{\Sigma}
\right)
\right]c_j(t/T),\nonumber\\
\end{eqnarray}
%%%%%%%%%%%%%%% Eq flavored SS %%%%%%%%%%%%%%%%
\begin{eqnarray}
\label{eq:Smult}
\int d^3 x \langle S^a(x)S^b(0)\rangle 
%\nonumber\\
%&&\hspace{-1.4in}
&\equiv& 
\frac{\delta_{ab}}{2}\int d^3 x \langle \bar{q}_{v_1}q_{v_1}(x)\bar{q}_{v_1}q_{v_1}(0)\rangle_c
\nonumber\\
&&\hspace{-1.5in}
=\frac{\delta_{ab}}{2}\left[
L^3\Sigma^2\left(\frac{x_1^{\rm eff}}{x_1}\right)^2
\frac{\partial_x\Sigma_{\nu}^{{\rm PQ}}(x_1^{\rm eff},\{z_i^{\rm eff}\})}{\Sigma}
-\frac{2\Sigma^2}{F^2}\frac{\nu^2}{x_1^2}
a(t/T)\right.\nonumber\\
&&\hspace{-1.4in}-\frac{2\Sigma^2}{F^2}
\frac{\Sigma_{\nu}^{{\rm PQ}}(x_1, \{z_i\})}{x_1\Sigma}b(t/T)\nonumber\\
&&\hspace{-1.4in}\left.+\frac{\Sigma^2}{2F^2}\sum_j^{N_f}\left[
\frac{4}{x_1^2-z_j^2}\left(
\frac{x_1\Sigma_{\nu}^{{\rm PQ}}(x_1, \{z_i\})}{\Sigma}
-\frac{z_j\Sigma_{\nu}^{{\rm full}(N_f, j)}(\{z_i\})}{\Sigma}
\right)
\right]c_j(t/T)\right],\nonumber\\
\end{eqnarray}
%%%%%%%%%%%%% Eq flavored PP %%%%%%%%%%%%%%%%%
\begin{eqnarray}
\label{eq:Pmult}
\int d^3 x \langle P^a(x)P^b(0)\rangle &\equiv&
-\frac{\delta_{ab}}{2}\int d^3 x\langle 
\bar{q}_{v_1}\gamma_5 q_{v_1}(x)\bar{q}_{v_1}\gamma_5 q_{v_1}(0)\rangle_c
\nonumber\\
&&\hspace{-1.5in}
=-\frac{\delta_{ab}}{2}\left[
-L^3\Sigma^2\left(\frac{x_1^{\rm eff}}{x_1}\right)^2
\frac{\Sigma_{\nu}^{{\rm PQ}}(x_1^{\rm eff},\{z_i^{\rm eff}\})}{x_1^{\rm eff}\Sigma}
+\frac{2\Sigma^2}{F^2}
\frac{\Delta\Sigma_{\nu}^{{\rm PQ}}(x_1,\{z_i\})}{\Sigma}a(t/T)
\right.\nonumber\\
&&\hspace{-1.4in}+\frac{2\Sigma^2}{F^2}\frac{\partial_x\Sigma_{\nu}^{{\rm PQ}}(x_1, \{z_i\})}{\Sigma}
b(t/T)\nonumber\\
&&\hspace{-1.4in}
\left.-\frac{\Sigma^2}{2F^2}\sum_j^{N_f}\left[
\frac{4}{x_1^2-z_j^2}\left(
\frac{x_1\Sigma_{\nu}^{{\rm PQ}}(x_1, \{z_i\})}{\Sigma}
-\frac{z_j\Sigma_{\nu}^{{\rm full}(N_f, j)}(\{z_i\})}{\Sigma}
\right)
\right]c_j(t/T)\right],\nonumber\\
\end{eqnarray}
where
$a(t/T),b(t/T)$ and $c_j(t/T)$ are defined by
%%%%%%%%%%% Eq  a, b, and c definition %%%%%%%%%%%
\begin{eqnarray}
a(t/T) &\equiv& \int d^3 x \bar{\Delta}(0|x)=\frac{T}{2}
\left[\left(\frac{t}{T}-\frac{1}{2}\right)^2-\frac{1}{12}\right],\\
b(t/T) &\equiv& \int d^3 x \bar{G}(0,0|x)
= \int d^3 x  \frac{1}{V}
\sum_{p\neq 0}
\frac{e^{ipx}}{p^4
\left(\sum^{N_f}_i \frac{1}{p^2+M^2_{ii}}\right)}
\nonumber\\
&=&\frac{1}{N_f}\frac{T}{2}
\left[\left(\frac{t}{T}-\frac{1}{2}\right)^2-\frac{1}{12}\right]
\nonumber\\
&&
-\left(\sum^{N_f}_i \frac{M^2_{ii}}{N_f^2}\right)\frac{T^3}{24}
\left[\left(\frac{t}{T}\right)^2\left(\frac{t}{T}-1\right)^2-\frac{1}{30}\right]
+O(M^4_{ii}),\\
c_j(t/T) &\equiv& \int d^3 x \bar{\Delta}(M^2_{jv}|x)=
\frac{\cosh (M_{jv}(T/2-t))}{2M_{jv}\sinh (M_{jv}T/2)}
-\frac{1}{M^2_{jv}T}\nonumber\\
&=&\frac{T}{2}\left[\left(\frac{t}{T}-\frac{1}{2}\right)^2-\frac{1}{12}\right]
+M^2_{jv}\frac{T^3}{24}
\left[\left(\frac{t}{T}\right)^2
\left(\frac{t}{T}-1\right)^2-\frac{1}{30}\right]
\nonumber\\&&+O(M^4_{jv}),
\end{eqnarray}
where $M^2_{jv} = m_j\Sigma /F^2$ and $M^2_{ii}=2m_i\Sigma/F^2$.

\begin{figure*}[pth]
  \includegraphics[width=14cm]{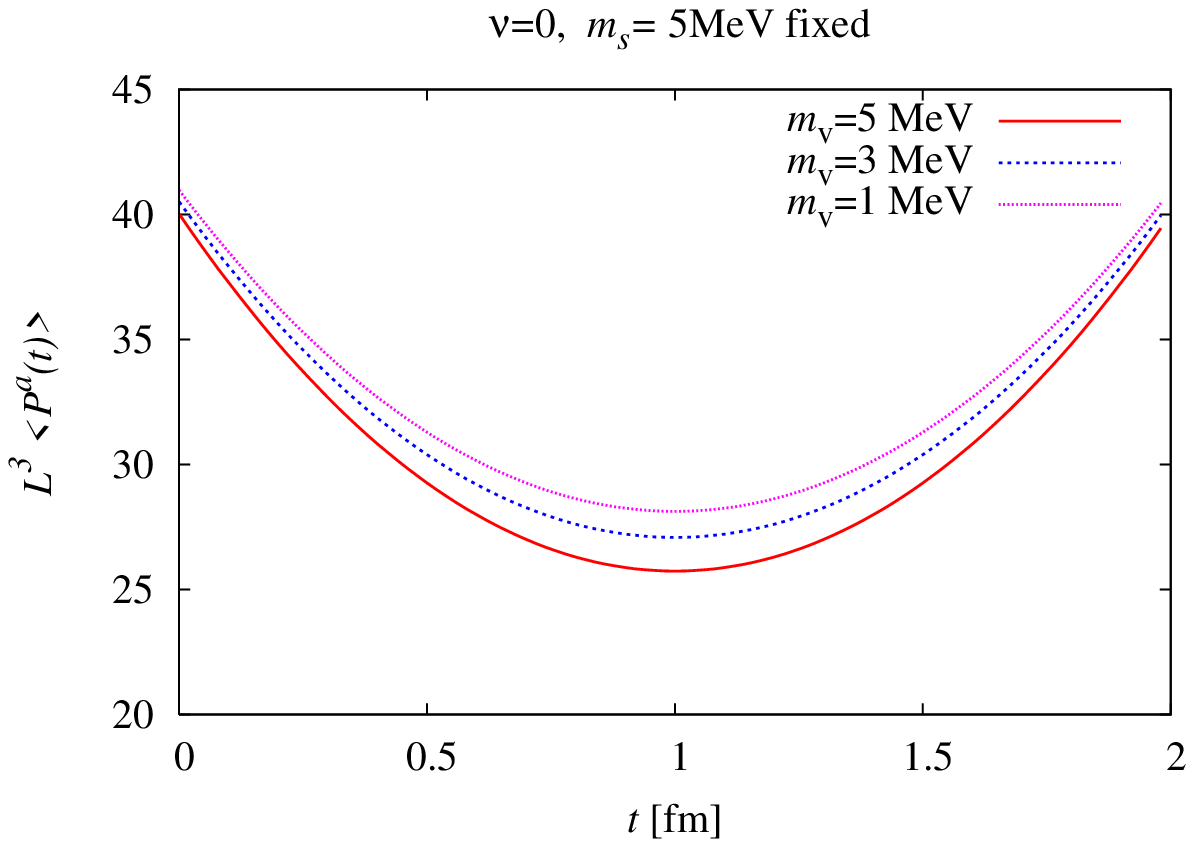}
  \includegraphics[width=14cm]{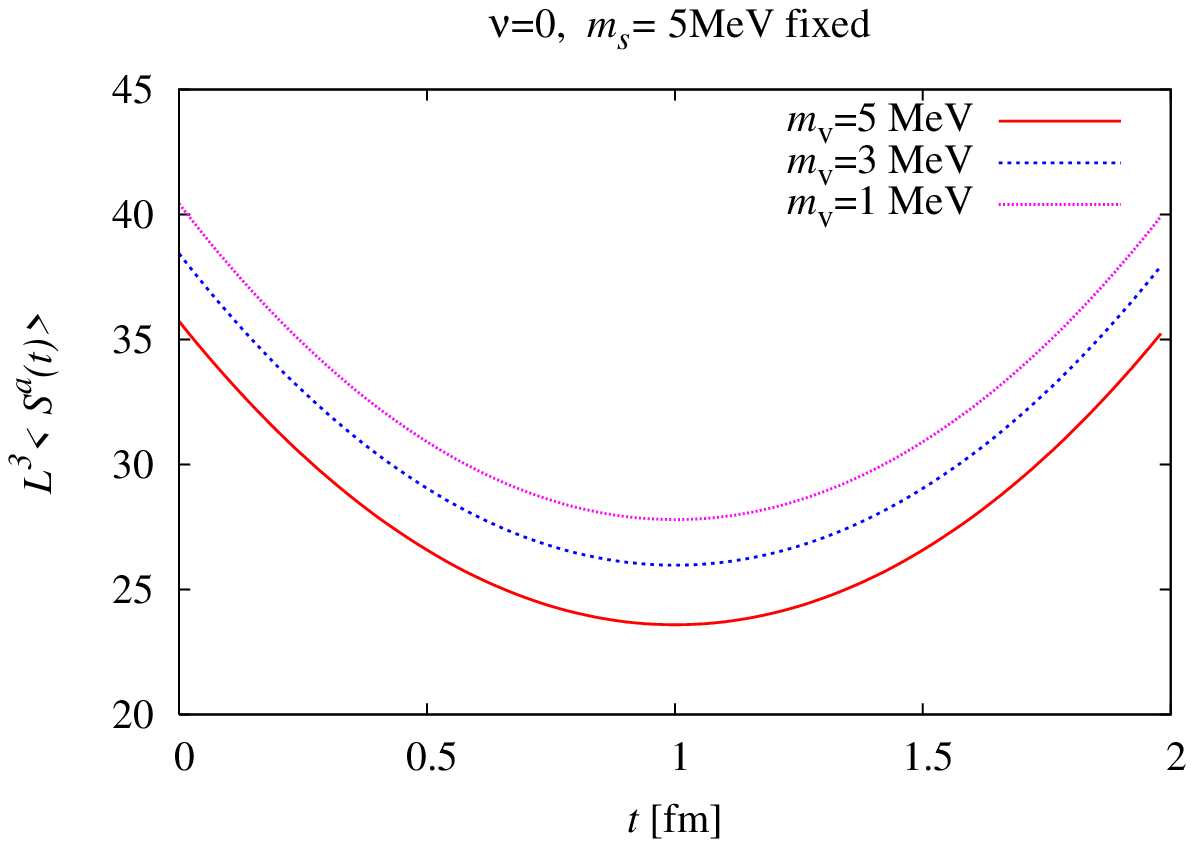}
  \caption{
    Pseudo-scalar (top) and scalar (bottom) flavored correlators
    with both valence and sea quarks in the $\epsilon$-regime, 
 here for $\nu=0$.
    The sea quark masses are fixed to 5 MeV.
    We use $\Sigma =$(250 MeV$)^3$, $F=$ 93 MeV, $x_1^{\rm eff}/x_1=1$
    and $L=T=$ 2 fm as inputs. 
  }
  \label{fig:prope-regime}
\end{figure*}

\begin{figure}[pth]
  \centering
  \includegraphics[width=14cm]{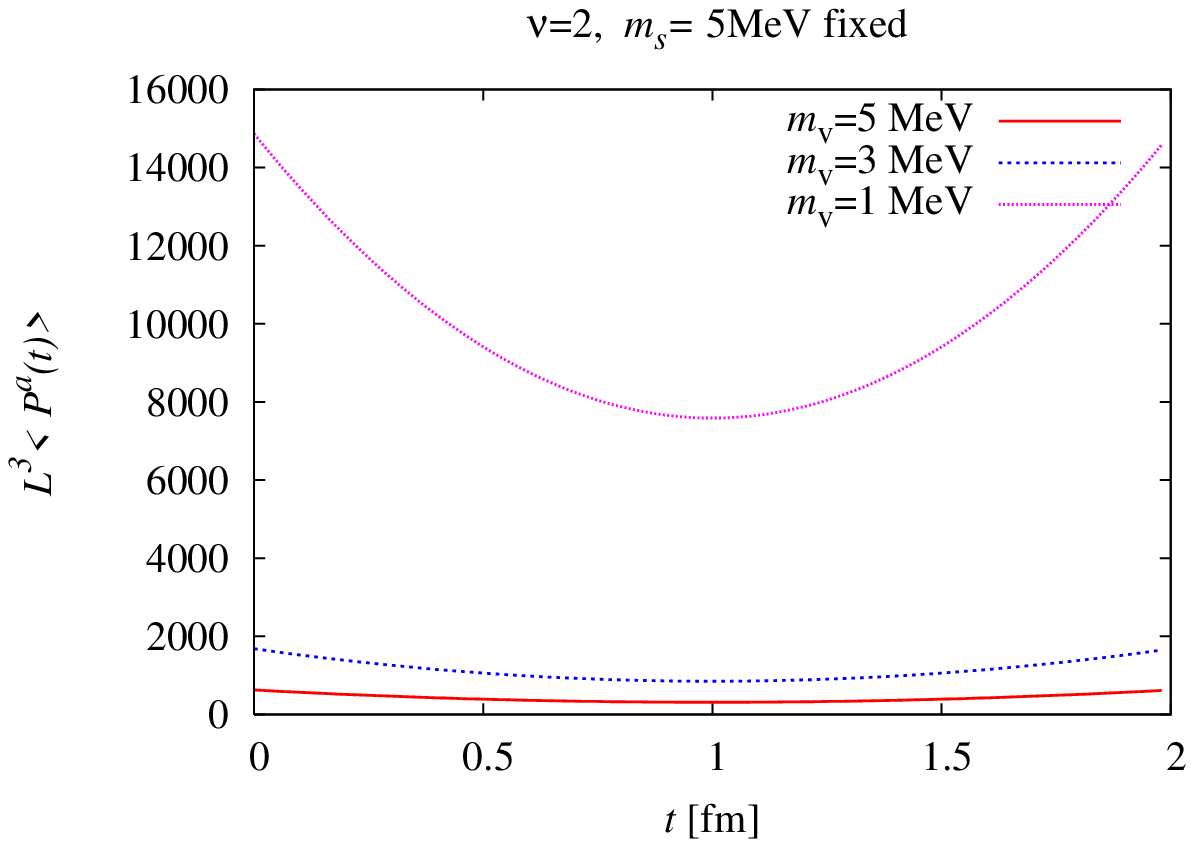}
  \includegraphics[width=14cm]{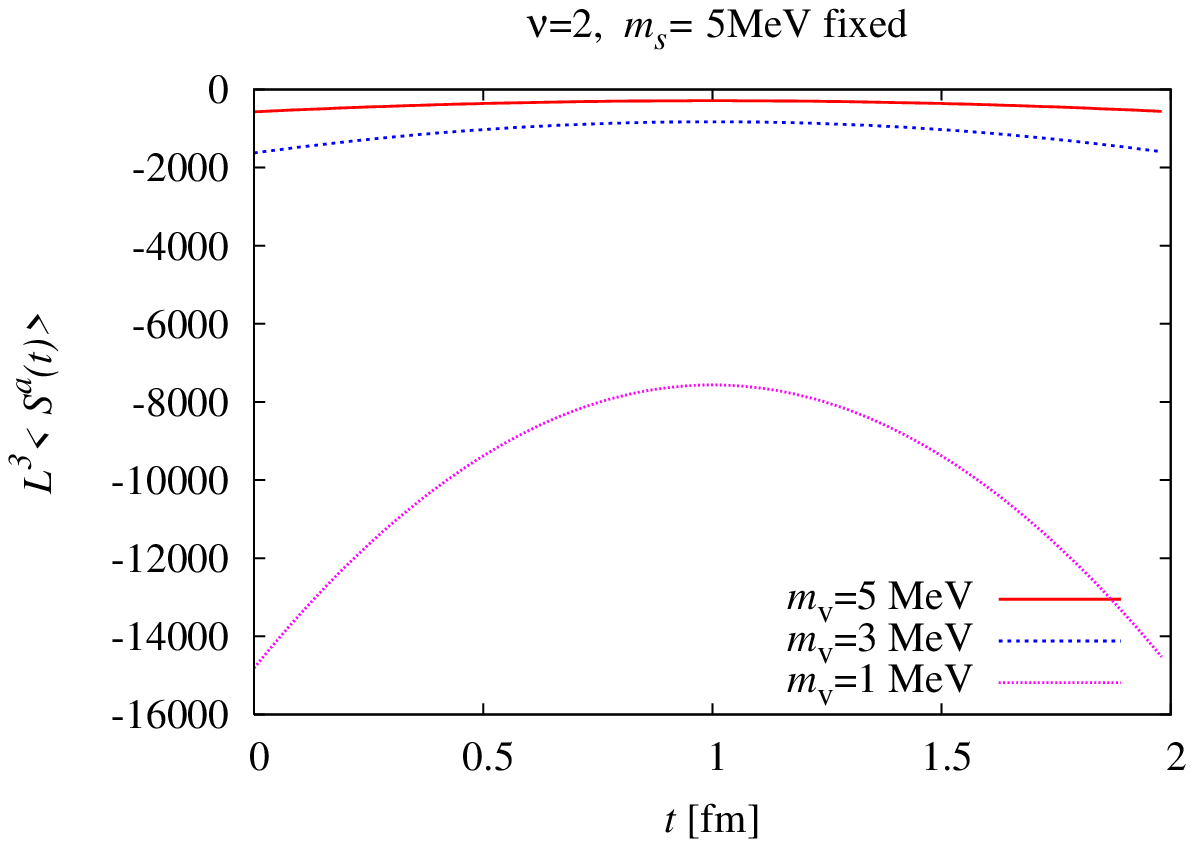}
  \caption{
    The same as Fig. \ref{fig:prope-regime} but with $\nu=2$.
    Note that the pseudo-scalar and scalar correlation functions
    almost sum up to zero. 
  }
  \label{fig:prope-regime2}
\end{figure}

\begin{figure}[pth]
  \centering
  \includegraphics[width=14cm]{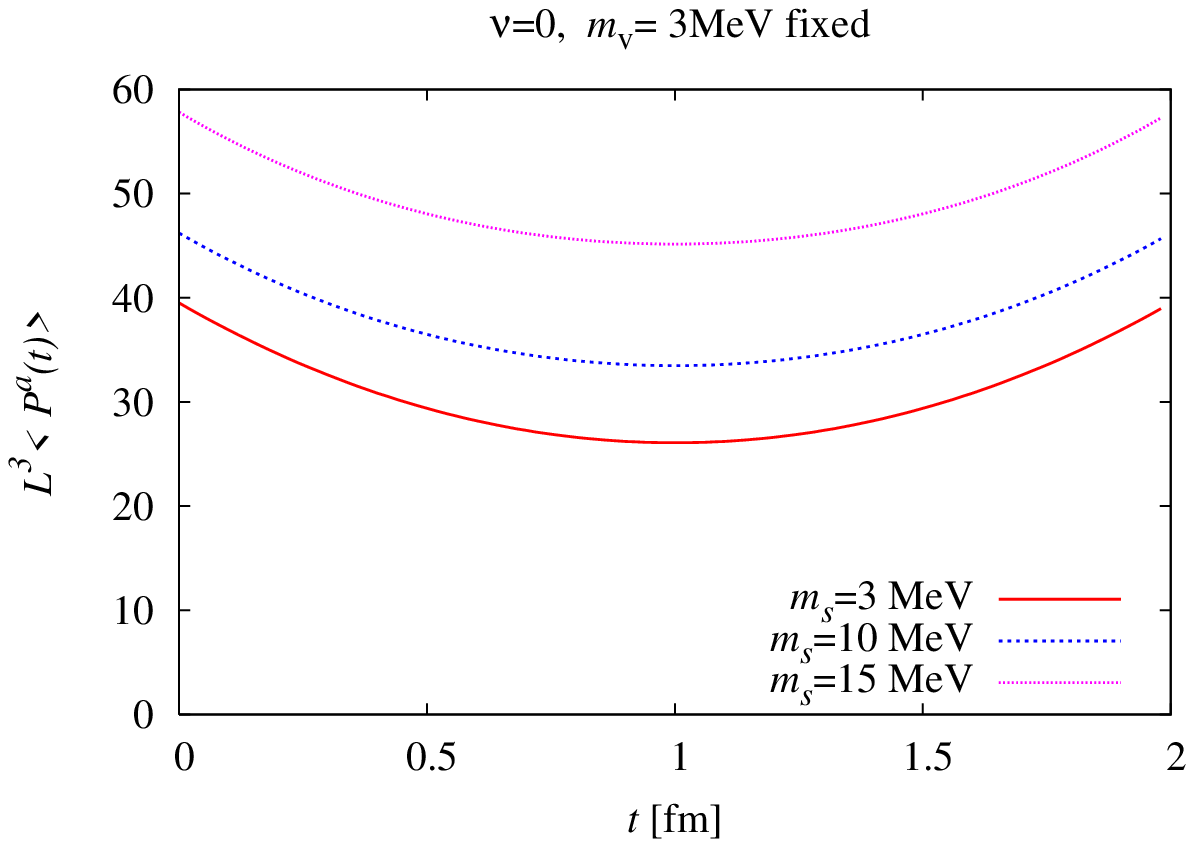}
  \includegraphics[width=14cm]{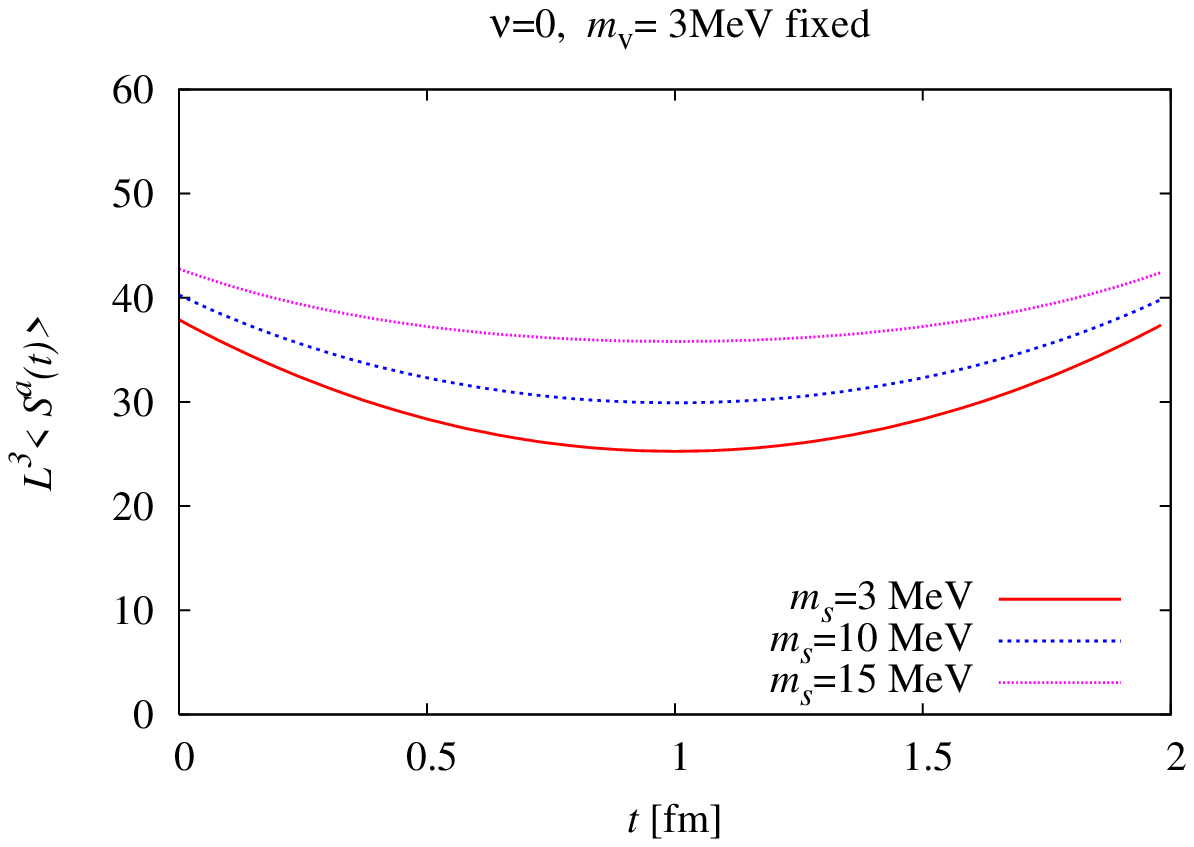}
  \caption{
    The sea quark mass dependence of
    pseudo-scalar (top) and scalar (bottom) flavored correlators
    with the valence quark mass fixed at $3$ MeV.
    The parameters are chosen the same as for Fig.\ref{fig:prope-regime}
    and Fig.\ref{fig:prope-regime2}.
     Here we assume that we can still ignore the $m_0$ and $\alpha$
 terms for sea quark masses up to $\sim 15$ MeV.
  }
  \label{fig:prophybrid}
\end{figure}

In Fig. \ref{fig:prope-regime}, Fig. \ref{fig:prope-regime2} and 
Fig. \ref{fig:prophybrid}, we plot, as examples, the flavored pseudo-scalar and
scalar correlators (we simply denote Eq.(\ref{eq:Smult}) 
and Eq.(\ref{eq:Pmult}) as $\langle S^a(t)\rangle$ and 
$\langle P^a(t)\rangle$ for $a=b$ ).
We use $\Sigma =$(250 MeV$)^3$, $F=$ 93 MeV, $x_1^{\rm eff}/x_1=1$
and $L=T=$ 2 fm as inputs.

%%%%%%%%%%%%%%%%%%%%%%%%%%%%%%%%%%%%%%%%%%%%%%%%%%%%%%%%%%%%
\subsection{Ward-Takahashi identities}
\label{sec:WTI}
%%%%%%%%%%%%%%%%%%%%%%%%%%%%%%%%%%%%%%%%%%%%%%%%%%%%%%%%%%%%

In this subsection, we check that the above results 
satisfy the Ward-Takahashi identities under the chiral rotation
of the degenerate $N_v$ valence quarks.
For the singlet chiral rotation, one obtains
\begin{equation}
\label{eq:WTIsinglet}
\langle (\partial_\mu A^0_\mu(x)-2mP^0(x)-2iN_v\omega(x){\cal O}(0)\rangle
=-\langle \delta{\cal O}(0)\rangle\delta(x),
\end{equation}
for any operator ${\cal O}(x)$, where 
\begin{eqnarray}
A^0_\mu(x)&=&\sum^{N_v}_{v=1}\bar{q}_v(x)\gamma_\mu\gamma_5q_v(x),\;\;\;
P^0(x)=\sum^{N_v}_{v=1}\bar{q}_v(x)\gamma_5q_v(x),
\nonumber\\
\omega(x)&=&\frac{1}{16\pi^2}\mbox{Tr}F_{\mu\nu}\tilde{F_{\mu\nu}}(x),
\end{eqnarray}
and $\delta {\cal O} $ denotes the chiral variation of ${\cal O}$.
Note that Eq.(\ref{eq:WTIsinglet}) holds not only in $\theta$-vacuum 
but also in a fixed topological sector.
The identities for ${\cal O}(x)=P^0(x)$ and ${\cal O}(x)=\omega(x)$
and their integration over the volume give an equation
\begin{equation}
\int d^4x \langle P^0(x)P^0(0)\rangle
= -\frac{N_v^2\nu^2}{m^2V} - \frac{\langle S^0(x)\rangle}{m}
= -\frac{N_v^2\nu^2}{m^2V} + 
\frac{N_v\Sigma^{{\rm PQ}}_{\nu}(x^{\rm eff}_1,\{z^{\rm eff}_i\})
x_1^{\rm eff}}{m x_1},
\end{equation}
where $S^0(x)=\sum^{N_v}_{v=1}\bar{q}_v(x)q_v(x),$ 
which coincides with Eq.(\ref{eq:Psinglet}). 
Note here we used 
\begin{eqnarray}
\int dt\; a(t/T) = \int dt\; b(t/T) = \int dt\; c_j(t/T) = 0.
\end{eqnarray}

In the same way, the flavored identity 
\begin{equation}
\label{eq:WTIflavored}
\langle (\partial_\mu A^a_\mu(x)-2mP^a(x)){\cal O}(0)\rangle
=-\langle \delta^a{\cal O}(0)\rangle\delta(x),
\end{equation}
for ${\cal O}(x)=P^a(x)$ and 
${\cal O}(x)=\partial_\mu A^a_\mu(x)$ gives
\begin{equation}
\int d^4x \langle P^a(x)P^a(0)\rangle
= -\frac{\langle S^0(x)\rangle}{2N_v m}=
\frac{\Sigma_{\nu}^{{\rm PQ}}(x_1^{\rm eff},\{z_i^{\rm eff}\})x_1^{\rm eff}}
{2m x_1},
\end{equation}
which is also consistent with Eq.(\ref{eq:Pmult}). 
Here the operator with the superscript ``$a$'' 
has the form of $\bar{q}_v(x)\gamma T^a q_v(x)$ 
with some gamma matrix $\gamma$, 
as in the conventional notation, where $T^a$ denotes
the $a$-th generator of $SU(N_v)$ group.

One can also confirm the similar identity \cite{Edwards:1998wx}
\begin{equation}
\int d^4x \langle S^a(x)S^a(0)\rangle
= \frac{1}{2N_v V}\frac{\partial}{\partial m_v}
\langle S^0(x)\rangle=\left(\frac{x_1^{\rm eff}}{x_1}\right)^2
\frac{\partial}{2\partial m_{v_1} }
\Sigma_{\nu}^{{\rm PQ}}(x_1^{\rm eff},\{z_i^{\rm eff}\}),
\end{equation}
is consistent with Eq.(\ref{eq:Smult}).
Since the asymptotic form  of the partially quenched 
condensate in the chiral limit is known \cite{Damgaard:1999ic}
\begin{equation}
\frac{\Sigma_{\nu}^{{\rm PQ}}(x_1,\{z_i\})}
{\Sigma }\sim \frac{|\nu|}{x_1} + {\cal O}(x_1),
\end{equation}
it is not difficult to see that the known ``quenched'' identities
for the $N_v=2$ case \cite{Edwards:1998wx}, 
\begin{eqnarray}
\int d^4x \left[\langle P^0(x)P^0(0)\rangle 
- 4 \langle S^a(x)S^a(0)\rangle\right]&\sim& 
\frac{4|\nu|}{m_v^2 V}-\frac{4\nu^2}{m_v^2 V},\\
\int d^4x \left[\langle P^a(x)P^a(0)\rangle 
- 4 \langle S^a(x)S^a(0)\rangle\right]&\sim& 
\frac{4|\nu|}{m_v^2 V},
\end{eqnarray}
also hold in the limit $m_v\to 0$ in this 
partially quenched theory.

%%%%%%%%%%%%%%%%%%%%%%%%%%%%%%%%%%%%%%%%%%%%%%%%%%%%%%%%%%%%%%
\subsection{Quenched and full degenerate $N_f$ flavor limits}
%%%%%%%%%%%%%%%%%%%%%%%%%%%%%%%%%%%%%%%%%%%%%%%%%%%%%%%%%%%%%%

In this subsection, we show how to reproduce the known results of 
both the fully quenched theory and the unquenched theory by taking
the $z_i\to \infty$ limit and the $x=z_1=z_2=\cdots = z$ limits,
respectively. The valence quarks are chosen to be degenerate
in the $\epsilon$-regime.\\

First consider the quenched limit of 
connected correlators Eq.(\ref{eq:Scon-deg}) 
and Eq.(\ref{eq:Pcon-deg}),
%%%%%%%%% Eq con SS quenched %%%%%%%%%%%%%%%%%
\begin{eqnarray}
\label{eq:Scon-que}
\langle S^a(x)S^a(0)\rangle^{que}&=&\lim_{z_i\to\infty}
\frac{1}{2}\langle \bar{q}_{v_1}q_{v_1}(x)\bar{q}_{v_1}q_{v_1}(0)\rangle_c
%\xrightarrow[{z_i\to \infty}]{}
%\hspace{1.3in}
\nonumber\\
&=&
\Sigma^2\left(\frac{x_1^{\rm eff}}{x_1}\right)^2
\frac{\partial_x\Sigma_{\nu}^{que}(x_1^{\rm eff})}{2\Sigma}
\nonumber\\
&&
+\frac{\Sigma^2}{2F^2}\left[ -\frac{2\nu^2}{x_1^2}
\bar{\Delta}(0|x)
%\nonumber\\
%&&
%-\frac{2\Sigma^2}{F^2}
-\frac{2\Sigma_{\nu}^{que}(x_1)}{x_1\Sigma}\bar{G}(0,0|x)
\right],
\nonumber\\
\end{eqnarray}
%%%%%%%%%%% Eq con PP quenched %%%%%%%%%%%%%%%%%%%%%
\begin{eqnarray}
\label{eq:Pcon-que}
\langle P^a(x)P^a(0)\rangle^{que}&=&\lim_{z_i\to\infty}
-\frac{1}{2}\langle \bar{q}_{v_1}\gamma_5 
q_{v_1}(x)\bar{q}_{v_1}\gamma_5 q_{v_1}(0)\rangle_c
%\xrightarrow[{z_i\to \infty}]{}\hspace{1.3in}
\nonumber\\
&=&\Sigma^2\left(\frac{x_1^{\rm eff}}{x_1}\right)^2
\frac{\Sigma_{\nu}^{que}(x_1^{\rm eff})}{2x_1^{\rm eff}\Sigma}
\nonumber\\
&&-\frac{\Sigma^2}{2F^2}\left[-2\left(1+\frac{\nu^2}{x_1^2}\right)
%-\frac{2\Delta\Sigma_{\nu}^{{\rm PQ}}(x_1,\{z_i\})}{\Sigma}
\bar{\Delta}(0|x)
+\frac{2\partial_x\Sigma_{\nu}^{que}(x_1)}{\Sigma}
\bar{G}(0,0|x)\right].\nonumber\\
\end{eqnarray}
Noting 
%%%%%%%%%%% Eq Delta, G, quenched limit %%%%%%%%%%%
\begin{eqnarray}
\bar{\Delta}(0|x) = \frac{1}{V}\sum_{p\neq 0}\frac{e^{ipx}}{p^2}, \;\;\;
\bar{G}(0,0|x) = \frac{1}{V}\sum_{p\neq 0}
\frac{1}{N_c}\left(\frac{e^{ipx}m_0^2}{p^4}+\frac{e^{ipx}\alpha}{p^2}\right),
\end{eqnarray}
in the quenched limit, one can see that 
Eq.(\ref{eq:Scon-que}) and Eq.(\ref{eq:Pcon-que}) agree with the
quenched results in \cite{Damgaard:2001js}.

Next we construct $N_v=1$ singlet correlation functions
in the quenched limit,
%%%%%%%%%%%% Eq singlet SS quenched %%%%%%%%%%%%%%
\begin{eqnarray}
\label{eq:Ssin-que}
\langle S^0(x)S^0(0)\rangle^{que}&=&\lim_{z_i\to\infty}(
%\nonumber\\
\langle \bar{q}_{v_1}q_{v_1}(x)\bar{q}_{v_1}q_{v_1}(0)\rangle_c
+\langle \bar{q}_{v_1}q_{v_1}(x)\bar{q}_{v_1}q_{v_1}(0)\rangle_d)
\nonumber\\
&&\hspace{-0.5in}=
%\xrightarrow[{z_i\to \infty}]{}
%(\Sigma^v_{\rm eff})^2
\Sigma^2\left(\frac{x_1^{\rm eff}}{x_1}\right)^2
\left[
\frac{\partial_x\Sigma_{\nu}^{que}(x_1^{\rm eff})}{\Sigma}
+1+\frac{\nu^2}{(x_1^{\rm eff})^2}
\right]
\nonumber\\
&&%\hspace{-1.5in}
+\frac{\Sigma^2}{2F^2}\left[ \left(-\frac{4\nu^2}{x_1^2}
+\frac{2\Sigma_{\nu}^{que}(x_1)}{x_1\Sigma}
-\frac{2\partial_x\Sigma_{\nu}^{que}(x_1)}{\Sigma}
\right)
\bar{\Delta}(0|x)\right.
\nonumber\\
&&-\left.\left(\frac{4\Sigma_{\nu}^{que}(x_1)}{x_1\Sigma}
-\frac{4\nu^2}{x_1^2}
\right)\bar{G}(0,0|x)
\right],
\end{eqnarray}
%%%%%%%%%%%% Eq singlet PP quenched %%%%%%%%%%%%%%
\begin{eqnarray}
\label{eq:Psin-que}
\langle P^0(x)P^0(0)\rangle^{que} \nonumber\\
&&\hspace{-0.5in}= \lim_{z_i\to\infty}
(\langle \bar{q}_{v_1}\gamma_5 
q_{v_1}(x)\bar{q}_{v_1}\gamma_5 q_{v_1}(0)\rangle_c
-\langle \bar{q}_{v_1}\gamma_5 
q_{v_1}(x)\bar{q}_{v_1}\gamma_5 q_{v_1}(0)\rangle_d)
\nonumber\\
&&%\hspace{-4in}
\hspace{-0.5in}=
%\xrightarrow[{z_i\to \infty}]{}
%(\Sigma^v_{\rm eff})^2
\Sigma^2\left(\frac{x_1^{\rm eff}}{x_1}\right)^2
\left(
\frac{\Sigma_{\nu}^{que}(x_1^{\rm eff})}{x_1^{\rm eff}\Sigma}
-\frac{\nu^2}{(x_1^{\rm eff})^2}
\right)
\nonumber\\
&&%\hspace{-4in}
-\frac{\Sigma^2}{2F^2}\left[-\left(4+\frac{4\nu^2}{x_1^2}
-\frac{2\Sigma_{\nu}^{que}(x_1)}{x_1\Sigma}
+\frac{2\partial_x\Sigma_{\nu}^{que}(x_1)}{\Sigma}
\right)\right.
%-\frac{2\Delta\Sigma_{\nu}^{{\rm PQ}}(x_1,\{z_i\})}{\Sigma}
\bar{\Delta}(0|x)
\nonumber\\
&&\left.
+\left(
\frac{4\partial_x\Sigma_{\nu}^{que}(x_1)}{\Sigma}
+4+\frac{4\nu^2}{x_1^2}
\right)
\bar{G}(0,0|x)\right],
\end{eqnarray}
which are also equivalent to the results of \cite{Damgaard:2001js}.

The full degenerate $N_f$-flavor limit ($x\to z =z_1=z_2=\cdots$)
of the connected correlators in the $\epsilon$-regime are
%%%%%%%%%%%% Eq flavored SS full %%%%%%%%%%%%%%
\begin{eqnarray}
\label{eq:Scon-full}
\langle S^a(x)S^a(0)\rangle^{full}&=&
\frac{1}{2}\langle \bar{q}_{v_1}q_{v_1}(x)\bar{q}_{v_1}q_{v_1}(0)\rangle_c
|_{x=z_1=z_2\cdots =z}
\nonumber\\
&&\hspace{-1in}=
%(\Sigma_{\rm eff})^2
\Sigma^2\left(
\left.\frac{z^{\rm eff}}{z}\right)^2
\frac{\partial_x\Sigma_{\nu}^{{\rm PQ}}(x,
\{z^{\rm eff},z^{\rm eff},\cdots \})}{2\Sigma}\right|_{x=z^{\rm eff}}
\nonumber\\
&&\hspace{-0.2in}-\frac{\Sigma^2}{2F^2}\left[
\frac{2\nu^2}{z^2}
+\frac{2\Sigma_{\nu}^{N_f}(z)}{N_fz\Sigma}
-N_f\left(
\left.\frac{\partial_x\Sigma_{\nu}^{{\rm PQ}}
(x, \{z,z,\cdots\})}{\Sigma}\right|_{x=z}\right.\right.
\nonumber\\
&&
%\hspace{-0.2in}
\left.\left.
+\frac{\Sigma_{\nu}^{N_f}(z)}{z\Sigma}
\right)\right]\bar{\Delta}(0|x),\nonumber\\
\end{eqnarray}
%%%%%%%%%%%% Eq flavored PP full %%%%%%%%%%%%%%
\begin{eqnarray}
\label{eq:Pcon-full}
\langle P^a(x)P^a(0)\rangle^{full}&=&
-\frac{1}{2}\langle \bar{q}_{v_1}\gamma_5 q_{v_1}(x)
\bar{q}_{v_1}\gamma_5 q_{v_1}(0)\rangle_c|_{x=z_1=z_2\cdots =z}
%(\Sigma_{\rm eff})^2
\nonumber\\
&&\hspace{-1in}
=
\Sigma^2\left(\frac{z^{\rm eff}}{z}\right)^2
\frac{\Sigma_{\nu}^{N_f}(z^{\rm eff})}{2z^{\rm eff}\Sigma}
\nonumber\\
&&\hspace{-.8in}
+\frac{\Sigma^2}{2F^2}\left[
-\frac{2\Delta\Sigma_{\nu}^{{\rm PQ}}(z,\{z,z,\cdots\})}{\Sigma}
%\right.
%\nonumber\\
%&&\hspace{-2in}
%\left.
-\left.\frac{2\partial_x\Sigma_{\nu}^{{\rm PQ}}(x, \{z,z,\cdots\})}{N_f\Sigma}
\right|_{x=z}
\right.
\nonumber\\
&&\hspace{-.8in}
\left.
+N_f\left(
\left.\frac{\partial_x\Sigma_{\nu}^{{\rm PQ}}
(x, \{z,z,\cdots\})}{\Sigma}\right|_{x=z}
+\frac{\Sigma_{\nu}^{N_f}(z)}{z\Sigma}
\right)\right]\bar{\Delta}(0|x).
\end{eqnarray}
Note here that we set $M^2_{iv}=M^2_{vv}=0$, 
$\bar{G}(0,0|x)=\bar{\Delta}(0|x)/N_f$, $\Sigma_{\nu}^{N_f}(z)/\Sigma$
is the full degenerate $N_f$-flavor condensate defined by 
Eq.(\ref{eq:Sigma-full-deg}),
and $z^{\rm eff}$ is given by Eq.(\ref{eq:sigmaeff-full-deg}).
%$\Sigma^v_{\rm eff} = \Sigma^i_{\rm eff}\equiv\Sigma_{\rm eff}$ is
%given by Eq.(\ref{eq:sigmaeff-full-deg}), and 
%$z^{\rm eff}=m\Sigma_{\rm eff} V$.
%{\bf change of notation.}
To eliminate the partially quenched expression, $\Sigma^{\rm PQ}/\Sigma$,
we use Eq.(\ref{eq:unitary}), Eq.(\ref{eq:sigmadif4}) and 
Eq.(\ref{eq:unitary-full}) to obtain
%%%%%%%%%% Eq flavored SS full 2 %%%%%%%%%%%%%%%%%%%%%
\begin{eqnarray}
\label{eq:Scon-full2}
%\frac{1}{2}\langle \bar{q}_{v_1}q_{v_1}(x)\bar{q}_{v_1}q_{v_1}(0)\rangle_c
\langle S^a(x)S^a(0)\rangle^{full}
=
\frac{N_f\Sigma^2}{2(N_f^2-1)}
\left(\frac{z^{\rm eff}}{z}\right)^2
%\frac{(\Sigma_{\rm eff})^2N_f}{2}
%\frac{\partial_x\Sigma_{\nu}^{{\rm PQ}}(z^{\rm eff},
%\{z^{\rm eff},z^{\rm eff},\cdots \})}{2\Sigma}
\left[
-\left(\frac{\Sigma^{N_f}_\nu(z^{\rm eff})}{\Sigma}\right)^2
\right.
\nonumber\\
\left.
-\frac{1}{N_f}\left(\frac{\Sigma^{N_f}_\nu(z^{\rm eff})}{\Sigma}\right)^\prime
-N_f\frac{\Sigma^{N_f}_\nu(z^{\rm eff})}{z^{\rm eff}\Sigma}
+1+\frac{\nu^2}{(z^{\rm eff})^2}
\right]
\nonumber\\
-\frac{\Sigma^2}{2(N_f^2-1)F^2}
\left[
(N_f^2-2)\frac{\nu^2}{z^2}
+\frac{(3N_f^2-2)}{N_f}\frac{\Sigma_{\nu}^{N_f}(z)}{z\Sigma}
\right.
\nonumber\\
%\hspace{-1.5in}
\left.
+N_f^2\left(
\left(\frac{\Sigma^{N_f}_\nu(z)}{\Sigma}\right)^2
+\frac{1}{N_f}\left(\frac{\Sigma^{N_f}_\nu(z)}{\Sigma}\right)^\prime
-1\right)\right]\bar{\Delta}(0|x),
\end{eqnarray}
%%%%%%%%%% Eq flavored PP full 2 %%%%%%%%%%%%%%%%%%%%%
\begin{eqnarray}
\label{eq:Pcon-full2}
%-\frac{1}{2}\langle \bar{q}_{v_1}\gamma_5 q_{v_1}(x)
%\bar{q}_{v_1}\gamma_5 q_{v_1}(0)\rangle_c
\langle P^a(x)P^a(0)\rangle^{full}
&=&
%(\Sigma_{\rm eff})^2
\Sigma^2\left(\frac{z^{\rm eff}}{z}\right)^2
\frac{\Sigma_{\nu}^{N_f}(z^{\rm eff})}{2z^{\rm eff}\Sigma}
\nonumber\\
&&
\hspace{-.5in}
+\frac{\Sigma^2}{2(N_f^2-1)F^2}\left[
(N_f^2-4)\frac{\nu^2}{z^2}+3N_f\frac{\Sigma_{\nu}^{N_f}(z)}{z\Sigma}
+N_f^2-4\right.
\nonumber\\
&&\left.
\hspace{-.5in}+(N_f^2+2)\left(
\left(\frac{\Sigma^{N_f}_\nu(z)}{\Sigma}\right)^2
+\frac{1}{N_f}\left(\frac{\Sigma^{N_f}_\nu(z)}{\Sigma}\right)^\prime
\right)\right]\bar{\Delta}(0|x),
\end{eqnarray}
which agree with those in the full $N_f$-flavor theory.

In the same way, one can see that the singlet correlators
%%%%%%%%%% Eq singlet SS full  %%%%%%%%%%%%%%%%%%%%%
\begin{eqnarray}
\label{eq:Sdis-full}
\langle S^0(x)S^0(0)\rangle^{full}
&=&
N_f\langle \bar{q}_{v_1}q_{v_1}(x)\bar{q}_{v_1}q_{v_1}(0)\rangle_c
+N_f^2\langle \bar{q}_{v_1}q_{v_1}(x)\bar{q}_{v_1}q_{v_1}(0)\rangle_d
\nonumber\\
&=&
\Sigma^2\left(\frac{z^{\rm eff}}{z}\right)^2
%(\Sigma_{\rm eff})^2
\left[
N_f\left.\frac{\partial_x\Sigma_{\nu}^{{\rm PQ}}(x,
\{z^{\rm eff},z^{\rm eff},\cdots \})}{\Sigma}\right|_{x=z^{\rm eff}}
\right.
\nonumber\\
&&\left.
-N_f^2\frac{\Delta\Sigma_{\nu}^{{\rm PQ}}
(z^{\rm eff},\{z^{\rm eff},z^{\rm eff},\cdots\})}{\Sigma}
\right]
\nonumber\\
&&+\frac{\Sigma^2}{2F^2}\left[
4(N_f^2-1)\frac{\Sigma_{\nu}^{N_f}(z)}{z\Sigma}
\right]\bar{\Delta}(0|x),\nonumber\\
\end{eqnarray}
%%%%%%%%%% Eq singlet PP full  %%%%%%%%%%%%%%%%%%%%%
\begin{eqnarray}
\label{eq:Pdis-full}
\langle P^0(x)P^0(0)\rangle^{full}
&=&
-N_f\langle \bar{q}_{v_1}\gamma_5 q_{v_1}(x)
\bar{q}_{v_1}\gamma_5 q_{v_1}(0)\rangle_c
-N^2_f\langle \bar{q}_{v_1}\gamma_5 q_{v_1}(x)
\bar{q}_{v_1}\gamma_5 q_{v_1}(0)\rangle_d
\nonumber\\
&=&%\hspace{-3in}
%(\Sigma_{\rm eff})^2
\Sigma^2\left(\frac{z^{\rm eff}}{z}\right)^2
\left[
N_f\frac{\Sigma_{\nu}^{N_f}(z^{\rm eff})}{z^{\rm eff}\Sigma}
-N_f^2\frac{\nu^2}{(z^{\rm eff})^2}\right]
%\hspace{0.5in}
\nonumber\\
&&%\hspace{-3in}
+\frac{\Sigma^2}{2F^2}\left[
4(N_f^2-1)
\left.\frac{\partial_x\Sigma_{\nu}^{{\rm PQ}}(x, \{z,z,\cdots\})}{\Sigma}
\right|_{x=z}
\right]\bar{\Delta}(0|x),\nonumber\\
\end{eqnarray}
are also consistent with the known expressions in \cite{Damgaard:2001js}.

%%%%%%%%%%%%%%%%%%%%%%%%%%%%%%%%%%%%%%%%%%%%%%
%%%%%%%%%%%%%%%%%%%%%%%%%%%%%%%%%%%%%%%%%%%%%%
\section{Conclusions}
\label{sec:conclusion}
\setcounter{equation}{0}
%%%%%%%%%%%%%%%%%%%%%%%%%%%%%%%%%%%%%%%%%%%%%%
%%%%%%%%%%%%%%%%%%%%%%%%%%%%%%%%%%%%%%%%%%%%%%

In this paper, we have discussed partially quenched
chiral perturbation theory (PQChPT) in the $\epsilon$-regime,
and in the mixed $\epsilon$ and $p$-regime.
%Based on the exact expression of the zero-mode partition function
%for $N_f+1$ fermionic quarks and one bosonic quark,
%a general analytic formula for the chiral condensate 
%of non-degenerate valence and sea quarks is obtained.
%We found that it converges to the known expressions
%both in the quenched limit and in the full theory limit.
%It was also found that the valence quarks receive 
%the different one loop corrections from that of sea quarks.

Using the 1-loop improved chiral condensate and its 
derivative as building blocks,
we have calculated various zero-mode group integrals 
in the replica limit. These integrals are necessary for the 
computation of mesonic correlation functions in the
partially quenched theory.
We have also derived a non-trivial identity which is 
a consequence of
unitarity of the graded or replicated group.

With these zero-mode integrals and the Feynman
rules for the non-zero modes,
we have calculated the mesonic correlation functions 
for both connected and disconnected 
pseudo-scalar and scalar channels with non-degenerate 
quark masses, both of the valence and sea kind.
Among others,
our results can be applied to the mesons that 
consist of two non-degenerate valence quarks.
For a demonstration, we have plotted the flavored 
pseudo-scalar and scalar 
correlators with a realistic choice of input parameters.
As expected, they
show a non-trivial valence (sea) quark mass dependence
with a fixed sea (valence) quark mass.

These meson correlators were shown to have
the correct quenched and degenerate 
full $N_f$-flavor theory limits.
We have not addressed the implicit flavor 
dependence of $\Sigma$, or $F$, the fundamental parameters
in the infinite volume limit.
In order to complete the smooth connection among
the theories with different number of flavors,
one has to match the value of them
\cite{Giusti:2004an, Hernandez:2006kz, DeGrand:2006uy, 
DeGrand:2006nv, Hasenfratz:2007yj}
The flavor dependence is expected to be rather weak,
%%%%%%%%%%% Eq matching %%%%%%%%%%%%%%%%%%%%%%%%%
%\begin{eqnarray}
%\Sigma \;\;\; (\mbox{\rm in $N_f-1$-flavor theory})
%&=& \Sigma^\prime \;\;\;(\mbox{\rm in $N_f$-flavor theory}),\\
%F \;\;\; (\mbox{\rm in $N_f-1$-flavor theory})
%&=& F^\prime \;\;\;(\mbox{\rm in $N_f$-flavor theory}),
%\end{eqnarray}
%where $\Sigma^\prime$ and $F^\prime$ involve 
%the accumulated loop corrections of the $N_f$-th quark 
%on the way to its infinite mass limit.
but the matching is interesting and important 
for the future works.

Our results are useful for the analysis of
unquenched lattice QCD simulations in many ways.
The various valence quark masses can be used
for each set of fixed sea quark masses.
%Even for the simulations with rather heavy sea quark masses,
%(perhaps even out of the $\epsilon$-regime)
Even if the physical pions are just barely in the
$\epsilon$-regime, 
one can put the valence pions very safely in
the $\epsilon$-regime and compare numerical 
data with our formulae for the partially quenched
correlation functions. 

%%%%%%%%%%%%%%%%%%%%%%%%%%%%%%%%%
\section*{Acknowledgments}
%%%%%%%%%%%%%%%%%%%%%%%%%%%%%%%%%
We would like to thank J.J.M.Verbaarschot and K.Splittorff
for fruitful discussions.
HF wish to thank the members of Niels Bohr Institute
for warm hospitality during his stay.
The work of HF was supported by a Grant-in-Aid of the Japanese
Ministry of Education (No.18840045), and the work of PHD
was supported in part by the European Community Network
ENRAGE (MRTN-CT-2004-005616).

%%%%%%%%%%%%%%% APPENDIX %%%%%%%%%%%%%%%%%%%
%%%%%%%%%%%%%%%%%%%%%%%%%%%%%%%%%%%%%%%%%%%%
\appendix
%%%%%%%%%%%%%%%%%%%%%%%%%%%%%%%%%%%%%%%%%%%%
%%%%%%%%%%%%%%%%%%%%%%%%%%%%%%%%%%%%%%%%%%%%
\section{Summary of group integrals}
\label{app:Uint}
\setcounter{equation}{0}
%%%%%%%%%%%%%%%%%%%%%%%%%%%%%%%%%%%%%%%%%%%%
%%%%%%%%%%%%%%%%%%%%%%%%%%%%%%%%%%%%%%%%%%%%

Here we summarize the group integrals in the replica limit,
which are necessary for the meson correlators. 
See Sec.\ref{sec:Uint} for the details.
The formulae for one valence index are,
%%%%%%%%%%%% 1-valence %%%%%%%%%%%%%%%%%
%%%%%%%%%%%%%%%%%%%%%%%%%%%%%%%%%%%%%%%%%
%%%%%%%%%%%%%%%%%%%%%%%%%%%%%%%%%%%%%%%%%
\begin{eqnarray}
%%%%%%%%%%%% Eq (U+U) %%%%%%%%%%%%%%%%% 
\frac{1}{2}\langle (U_{v v}+U^\dagger_{v v}) \rangle
&=&
\frac{\Sigma^{{\rm PQ}}_{\nu}(x,\{z_i\})}{\Sigma},\\
%%%%%%%%%%%% Eq (U+U)^2 %%%%%%%%%%%%%%%%% 
%\lim_{N_v\to 0}\frac{1}{N_v}\sum_v
\frac{1}{4}\langle 
(U_{v v}+U_{v v}^\dagger)^2 \rangle
&=&
\frac{\partial_x\Sigma^{{\rm PQ}}_{\nu}(x,\{z_i\})}{\Sigma}
-\frac{\Delta \Sigma^{{\rm PQ}}_{\nu}(x,\{z_i\})}{\Sigma},\\
%%%%%%%%%%%% Eq (U-U) %%%%%%%%%%%%%%%%% 
%\lim_{N_v\to 0}\frac{1}{N_v}\sum_v
\frac{1}{2}\langle (U_{v v}-U^\dagger_{v v}) \rangle
&=& -\frac{\nu}{x},\\
%%%%%%%%%%%% Eq (U-U)^2 %%%%%%%%%%%%%%%%% 
%\label{eq:diag-}
%\lim_{N_v\to 0}\frac{1}{N_v}\sum_v
\frac{1}{4}\langle 
(U_{v v}-U_{v v}^\dagger)^2 \rangle
&=& -\frac{\Sigma^{{\rm PQ}}_\nu(x,\{z_i\})}{x\Sigma} + \frac{\nu^2}{x^2},
%\\
%%%%%%%%%%%% Eq (U^2-U^2) %%%%%%%%%%%%%%%%% 
%\frac{1}{4}\langle (U_{v v})^2-(U_{v v}^\dagger)^2 \rangle
%&=&\frac{\nu}{x^2}-\frac{\nu\Sigma^{{\rm PQ}}_\nu(x,\{z_i\})}{x\Sigma},
\end{eqnarray}
\begin{eqnarray}
%%%%%%%%%%%% Eq UU  %%%%%%%%%%%
\langle U_{v v}U^\dagger_{v v} \rangle
&=& \frac{1}{4}\langle 
(U_{v v}+U_{v v}^\dagger)^2 \rangle
- \frac{1}{4}\langle 
(U_{v v}-U_{v v}^\dagger)^2 \rangle\nonumber\\
&=& \frac{\partial_x\Sigma^{{\rm PQ}}_{\nu}(x,\{z_i\})}{\Sigma}
-\frac{\Delta \Sigma^{{\rm PQ}}_{\nu}(x,\{z_i\})}{\Sigma}
+\frac{\Sigma^{{\rm PQ}}_\nu(x,\{z_i\})}{x\Sigma} - \frac{\nu^2}{x^2}.
\end{eqnarray}
For two valence indices,
%%%%%%%%%%%% 2-valences %%%%%%%%%%%%%%%%%
%%%%%%%%%%%%%%%%%%%%%%%%%%%%%%%%%%%%%%%%%
%%%%%%%%%%%%%%%%%%%%%%%%%%%%%%%%%%%%%%%%%
\begin{eqnarray}
%%%%%%%%%%% Eq (U+U)11(U+U)22 %%%%%%%%%%%%
\frac{1}{4}\langle 
(U_{v_1 v_1}+U_{v_1 v_1}^\dagger) (U_{v_2 v_2}+U_{v_2 v_2}^\dagger) \rangle
&=& D_\nu^{{\rm PQ}}(x_1,x_2,\{z_i\}),\\
%%%%%%%%%%% Eq (U-U)11(U-U)22 %%%%%%%%%%%%
\frac{1}{4}\langle 
(U_{v_1 v_1}-U_{v_1 v_1}^\dagger) (U_{v_2 v_2}-U_{v_2 v_2}^\dagger)\rangle
&=& \frac{\nu^2}{x_1x_2},\\
%%%%%%%%%%% Eq U11U22 %%%%%%%%%%%%
\langle U_{v_1 v_1}U_{v_2 v_2}\rangle + 
\langle U^\dagger_{v_1 v_1}U^\dagger_{v_2 v_2}\rangle
&=& 2D_\nu^{{\rm PQ}}(x_1,x_2,\{z_i\})+\frac{2\nu^2}{x_1x_2},
\end{eqnarray}
where $D_\nu^{{\rm PQ}}$ is defined in Eq.(\ref{eq:DPQ}). Similarly,
\begin{eqnarray}
%%%%%%%%%%% Eq (U+U)^2 = UU %%%%%%%%%%%%
\frac{1}{4}\langle 
(U_{v_1 v_2}\pm U_{v_2 v_1}^\dagger)^2\rangle
&=& \frac{1}{4}\langle (U_{v_2 v_1}\pm U_{v_1 v_2}^\dagger)^2\rangle
\nonumber\\
&=& \frac{\pm 1}{2}\langle U_{v_1 v_2}U_{v_2 v_1}^\dagger\rangle
= \frac{\pm 1}{2}\langle U_{v_2 v_1}U_{v_1 v_2}^\dagger\rangle
\nonumber\\
&=&
\frac{\pm 1}{x_1^2-x_2^2}
\left(x_1\frac{\Sigma_{\nu}^{{\rm PQ}}(x_1,\{z_i\})}{\Sigma}
-x_2\frac{\Sigma_{\nu}^{{\rm PQ}}(x_2,\{z_i\})}{\Sigma}\right),
\nonumber\\
\\
%%%%%%%%%%% Eq U^2 + U^2 =0  %%%%%%%%%%%%
\frac{1}{4}\langle 
U_{v_1 v_2}^2+ (U_{v_2 v_1}^\dagger)^2\rangle&=&0,
\end{eqnarray}
\begin{eqnarray}
%%%%%%%%%%% Eq (U+U)12(U+U)21 %%%%%%%%%%%%
\frac{1}{4}\langle 
(U_{v_1 v_2}\pm U_{v_2 v_1}^\dagger)(U_{v_2 v_1}\pm U_{v_1 v_2}^\dagger)\rangle
=\hspace{1.5in}\nonumber\\ 
\frac{1}{x_1^2-x_2^2}\left(x_2\frac{\Sigma_{\nu}^{{\rm PQ}}(x_1,\{z_i\})}{\Sigma}
-x_1\frac{\Sigma_{\nu}^{{\rm PQ}}(x_2,\{z_i\})}{\Sigma}\right),
\end{eqnarray}
are obtained.

\end{document}